\documentclass[usenatbib]{mn2e}
\usepackage{txfonts}
\usepackage{epsfig}
\usepackage{lscape}
\usepackage{graphicx}

\def\newline{\hfil\break}

\def\aj{{AJ}}                   
\def\araa{{ARA\&A}}             
\def\apj{{ApJ}}                 
\def\apjl{{ApJ}}                
\def\apjs{{ApJS}}               
\def\aap{{A\&A}}                
\def\mnras{{MNRAS}}             
\def\nat{{Nature}}              

\title[The Building Up of the Black Hole Mass - Stellar Mass Relation]{The Building Up of the Black Hole Mass - Stellar Mass Relation}
\author[Lamastra et al.]{A. Lamastra$^1$\thanks{E-mail: lamastra@mporzio.astro.it}, N. Menci$^1$, R. Maiolino$^1$, F. Fiore$^1$, A. Merloni$^2$\\
$^1$ INAF - Osservatorio Astronomico di Roma, via di Frascati
33, I-00040 Monteporzio, Italy\\
$^2$ Max-Planck-Institut f\"ur Extraterrestrische Physik, Giessenbachstrasse, D-85741 Garching, 
Germany 
}

\date{Submitted:}

\pagerange{\pageref{firstpage}--\pageref{lastpage}} \pubyear{}

\begin{document}

\maketitle

\begin{abstract}
We derive the growth of SMBHs relative to the stellar content of their host galaxy predicted under the assumption of 
BH accretion triggered by galaxy encounters occurring  during their merging histories.
The latter are described through Monte Carlo realizations, and are connected to gas processes, star formation and BH accretion using a semi-analytic model of galaxy formation in a cosmological framework. This allows us to connect the star formation process in the host galaxies to the growth of  Supermassive Black Holes. We show that, within this framework, the ratio 
 $\Gamma\equiv (M_{BH}/M_*)(z)/(M_{BH}/M_*)(z=0)$ between the Black Hole mass and the galactic stellar mass (normalized to the local value)
 depends on both BH mass and redshift. While the average value and the spread of $\Gamma(z)$ increase with $z$, such an effect is larger for massive BHs, reaching values $\Gamma\approx 5$ for massive Black Holes ($M\geq 10^9$ $M_{\odot}$) at $z\gtrsim 4$, in agreement with recent observations of high-redshift QSOs; this is due to the the effectiveness of interactions in triggering BH accretion in high-density environments (where massive haloes form) at high redshifts. 

To investigate how different observations of $\Gamma (z)$  fit within our framework,  
we worked out specific predictions for sub-samples of the simulated galaxies corresponding to the different observational samples for which  measurements of $\Gamma$ have been obtained. We found that for Broad Line AGNs at intermediate redshifts $1\lesssim z\lesssim 2$ values of $\Gamma\approx 2$ are expected, with a mild trend toward larger value for increasing BH mass.  Instead, when we select from our Monte Carlo simulations only extremely gas rich, rapidly star forming galaxies at the epoch of peak in the cosmic star formation ($2\leq z\leq 3$), we find low values $0.3\leq \Gamma\leq 1.5$, consistent with recent observational findings on samples of sub-mm galaxies; in the framework of our model, these objects end up at $z=0$ in low-to-intermediate mass BHs  ($M\leq 10^9$ $M_{\odot}$), and they do not represent typical paths leading to local massive galaxies. The latter have formed preferentially through paths (in the $M_*-M_{BH}$ plane) passing {\it above} the 
local  $M_*-M_{BH}$ relation. We discuss how the global picture emerging from the model 
is consistent with a downsizing scenario, where massive BHs accrete a larger fraction of their final mass at high redshifts $z\geq 4$

\end{abstract}


\section{Introduction}

The discovery of tight local correlations between the mass $M_{BH}$ of Supermassive Black Holes (SMBHs) and global properties of their host galaxies, like the stellar mass $M_*$ or the velocity dispersion $\sigma_*$,  (Kormendy \& Richstone 1995; Magorrian et al. 1998; Ho 1999; Gebhardt et al. 2000; Ferrarese \& Merritt 2000; Marconi \& Hunt 2003; H\"aring \& Rix 2004; Kormendy \& Bender 2009) constitutes an important breakthrough in the understanding of galaxy evolution. If the growth of SMBHs is considerably  contributed by accretion over cosmological times (Soltan 1982; Yu \& Tremaine 2002; Marconi et al. 2004; Merloni \& Heinz 2008) this implies that most, if not all, galaxies have hosted an Active Galactic Nucleus (AGN) in the past, and that a strong physical connection exists between galaxy formation and the growth of SMBHs. 

Understanding the mechanisms responsible for such a connection, and establishing the relative time scales for star formation/assembly  and for SMBHs growth, requires the measurement of the above correlations at higher redshifts. For such a scope, the most 
straightforward correlation to study is the $M_{BH}-M_*$ relation, since the
other fundamental relation  $M_{BH}-\sigma_*$ would require the measurement of
stellar velocity dispersions which are very difficult to achieve at $z\gtrsim
1$ in normal galaxies.

The $M_{BH}-M_*$ relation in the local Universe has been widely investigated in
the literature. Magorrian et al. (1998) were the first to find a correlation
between the black hole mass and the stellar mass,
however they found  a very large scatter ($\sim$0.51 dex). This relation was
then re-examined using more reliable $M_{BH}$ and $M_*$ measurements (Merritt
\& Ferrarese 2001, Marconi \& Hunt 2003, H\"aring \& Rix 2004);  these studies
showed that the scatter in the $M_{BH}-M_*$ relation is comparable to the
scatter in the $M_{BH}-\sigma_*$ relation ($\sim$0.3 dex, see H\"aring \& Rix
2004). It is also worth noting that the relation is BH mass dependent (H\"aring \& Rix
2004), although the mean value of the $M_{BH}/M_*$ ratio obtained from the H\"aring
\& Rix (2004) sample: $\langle log(M_{BH}/M_*)\rangle\simeq$-2.8,
is consistent with those derived by Merritt \& Ferrarese (2001) and Marconi \&
Hunt (2003) when the latter is lowered by a factor
of 5/3 to account for more recent estimates of the bulge virial masses (see Maiolino et al. 2007).\\

The SMBH masses at $z\gtrsim 0.5$  are generally
estimated in luminous type 1 AGNs like quasars (QSOs) and radio galaxies; for example, virial BH mass estimators are based on the assumption that the broad-line region of the AGNs
are dominated by the gravity of the SMBH (see, e.g.,  Peterson \& Wandel 2000), and provide mass estimates of SMBHs at higher redshifts (see  Wandel, Peterson \& Malkan 1999; Kaspi et al. 2000; McLure \& Dunlop 2002, Merloni et al. 2009), extending out to $z\approx 6$ (e.g., Willott, McLure \&  Jarvis 2003; McLure \& Dunlop 2004; Vestergaard et al. 2004; Walter et al. 2004; Riechers et al. 2008). 
The evolution of the $M_{BH}/M_*$ ratio from the local value is generally quantified in terms of the parameter:
\begin{equation}
\Gamma\equiv (M_{BH}/M_*)(z)/(M_{BH}/M_*)(z=0).
\end{equation}

Although the AGN activity makes it difficult to derive stellar masses for the
host galaxies at $z\gtrsim 1.5$, several observational works suggest that at
such redshifts the mean $M_{BH}/M_*$ ratio is significantly higher than the
present value, corresponding to $\Gamma>1$.\\

Peng et al. (2006), applying the virial method on a sample of 31
gravitationally lensed AGNs and 20 non-lensed AGN and deriving the host galaxy
stellar mass from the R-band luminosity, found the $M_{BH}/M_*$ ratio to be
higher than the present value by a factor $\Gamma \approx 2$ for AGNs at $1\leq z \leq 1.7$ and by a factor of $\Gamma\approx 4$ at $z\geq 1.7$; analysing the hosts of 89 broad-line AGNs in the COSMOS survey Merloni et al. (2009) derive a positive evolution fitted with $\Gamma\approx (1+z)^{0.74}$ in the redshift range $1\leq z\leq 2.2$; exploiting radio-loud unification  McLure et al. (2006) obtained virial (linewidth) BH mass estimates from the 3C RR quasars, and the  stellar mass estimates from the 3C RR radio galaxies, thereby providing black hole and stellar mass estimates for a single population of early-type galaxies, finding $\Gamma\approx (1+z)^2$ for $z\lesssim 2$; 
at even higher redshifts  $z\geq 4$, measurements of the dynamical mass from the CO emission of molecular gas in the hosts of 
high-redshift QSOs (Walter et al. 2004;  Riechers et al. 2008) yield large values of $\Gamma=5-10$. 

While the above studies indicate that the growth of SMBHs is faster than the stellar mass assembly, they all 
focus on luminous AGNs, and are thus biased towards selecting the most massive SMBHs (see Lauer et al. 2007). 
In contrast to the above results, from the analysis of a sample of sub-mm selected galaxies (SMGs) 
Borys et al. (2005) find $\Gamma$ to decrease with redshift for $z\lesssim 2$; more recently, Alexander et al. (2008) found $\Gamma(z=2)=0.3$ for SMGs at $z\approx 2$ , indicating that for such objects the growth of SMBHs actually lags that of the host stellar mass. These latter studies are affected by a different bias, in that they select ultra-luminous ($L\gtrsim 10^{12}\,L_{\odot}$), gas-rich galaxies (with gas fraction relative to baryonic mass $f_{gas}\gtrsim 0.6$, see Greve et al. 2005; Tacconi et al. 2008), characterized by the conversion of a significant fraction of their initial gas reservoir of $10^{10}-10^{11}$ $M_{\odot}$ into stars in a short times scale of a few $10^8$ yrs (see Tacconi 2008 and references therein);  
given the correlation between star formation rate and stellar mass, such observations tend to be biased towards massive stellar hosts. 

Attempts to derive the evolution of the global $\langle\Gamma (z)\rangle$ averaged over for \textit{whole} AGN population requires a different approach, based on integrated observables. 
By comparing the redshift evolution of the integrated BH and of the stellar mass densities Merloni, Rudnick \& Di Matteo (2004) found that $\langle \Gamma\rangle \approx (1+z)^{\alpha}$ with $\alpha \simeq$0.5. Such a results is consistent with the upper limits on the evolution of  $\langle \Gamma\rangle$ 
derived by Hopkins et al. (2006) for $z \lesssim $ 2.

Thus, a major challenge for next studies of AGN and galaxy formation is understanding which of the above classes of  
paths $\Gamma(z)$ is the dominant one in the cosmological evolution of galaxies and SMBHs, and what are the physical mechanisms determining the particular path followed by a galaxy as a function of its properties in different cosmic epochs.  
While on the observational side such a task  would require a large, unbiased sample of AGNs,  a complementary approach is to consider cosmological models of galaxy and SMBH evolution which assume a given physical mechanism for the AGN feeding and for its connections with the evolution of the host galaxy; then, comparing the predictions of such models for specific classes of objects with different existing observations would constitute a powerful probe for the proposed physical link between galaxies and SMBHs.  

Indeed, in the recent years, significants steps forward had been taken toward the modelling of the cosmological co-evolution of galaxies and AGNs, starting from the seminal papers by Silk \& Rees (1998); Fabian (1999); Cavaliere \& Vittorini (2000); Wyithe \& Loeb (2003). On the one hand, recent high-resolution hydrodynamic N-body simulations (see, e.g., Di Matteo, Springel \& Hernquist 2005, Springel 2005; Hopkins 2005, 2006) have shown the importance of galaxy mergers as triggers for AGN accretion, and the role of the AGN energy feedback in the subsequent evolution of the host galaxy. However, such simulations are affected by a limited exploration of the sub-grid prescriptions for the physics of SMBHs; most important, they necessarely focus on specific galaxy systems, making it difficult to assess the  statistical relevance of a particular path in the building up of the local $M_{BH}-M_*$ relation. On the other hand, semi-analytic models (SAMs, Monaco, Salucci \& Danese 2000, Kauffmann \& Haenhelt 2000; Volonteri, Haardt \& Madau 2003; Granato et al. 2004; Menci et al. 2006; Croton et al. 2006; Bower et al. 2006; Marulli et al. 2008) can probe the whole set of possible paths followed by galaxies in the $M_{BH}-M_*$ plane during their evolution. Since their results depend on the  mechanisms assumed for the trigger of the AGN accretion and for the AGN energy feedback, SAMs are suitable to probe the statistical outcomes provided by the implemented 
physical processes linking SMBHs with their host galaxy. 

Here we adopt a state-of-the-art SAM (Menci et al. 2006, 2008) which implements a specific, physical model for the BH accretion  based on interactions as triggers for both the fueling of SMBHs and the star bursts. In turn, the interaction rate, the amount of gas available for BH accretion, and the stellar content of galaxies are computed from Monte Carlo realizations of galaxy merging trees, enabling us to follow the evolutionary paths of galaxies and of their SMBHs over cosmological times, from the collapse of their progenitors from the primordial density field to the present. Based on such a model we derive the evolutionary paths for the growth of SMBHs relative to that of the stellar mass of their host galaxies, to investigate which are the dominant paths 
followed by different galaxy populations in building up the local $M_{BH}-M_*$ relation in a scenario characterized by AGN fueling 
driven by galaxy interactions. We then compare different model predictions for the $M_{BH}/M_*$ ratio of 
luminous intermediate- and high-redshift galaxies, and for intermediate redshift SMGs with existing observations, and discuss the physical origin of the different paths $\Gamma(z)$ followed by galaxies and SMBHs during their evolution. Finally, we shall provide 
specific predictions typical of the scenario characterized by interaction-driven AGN fueling, to be tested with next experiments.

\section{The Model}

We use the semi-analytic model as is described in details in Menci et al. (2005, 2006, 2008); this connects, within a cosmological framework,
the accretion onto SMBHs and the ensuing AGN activities with the  evolution of galaxies. Here we recall the basic points.
\subsection{Hierarchical Galaxy Evolution}

Galaxy formation and evolution is driven by the collapse and growth of dark
matter (DM) haloes, which originate by gravitational instability of  overdense
regions in the primordial DM density field. This is taken to be a random,
Gaussian  density field with Cold Dark Matter (CDM) power spectrum within the
''concordance cosmology" (Spergel et al. 2007) for which we adopt round
parameters  $\Omega_{\Lambda}=0.7$, $\Omega_{0}=0.3$, baryonic density
$\Omega_b=0.04$ and Hubble constant (in units of 100 km/s/Mpc) $h=0.7$. The
normalization of the spectrum is taken to be $\sigma_8=0.9$ in terms of the variance
of the field smoothed over regions of 8 $h^{-1}$ Mpc.

The merging rates of the DM haloes are provided by the Extended Press \& Schechter formalism (see
Bond et al. 1991; Lacey \& Cole 1993).  The clumps included into larger DM haloes
may survive as satellites, or merge to form larger galaxies due to binary
aggregations,  or coalesce into the central dominant galaxy due to dynamical
friction; these processes take place over time scales that grow longer over
cosmic time, so the number of satellite galaxies increases as the DM host haloes
grow from groups to clusters (see Menci et al. 2005, 2006).

The processes connecting the baryonic components to the growing DM 
haloes (with mass $m$) are described in our previous papers (e.g., Menci et al. 2005). 
The gas at virial equilibrium with the DM potential wells undergoes radiative cooling. 
The cooled gas mass $m_c$ settles into a rotationally supported disc with radius $r_d$
(typically ranging from $1$ to $5 $ kpc), rotation velocity $v_d$
and dynamical time $t_d=r_d/v_d$. Two channels of star formation 
may convert part of such a gas into stars: \newline
i) quiescent star formation, corresponding to the gradual conversion into
stars at a rate $\dot m_*\propto m_c/t_d$\newline
ii) starbursts triggered by interactions, which destabilize part of the cold gas available by inducing
loss of angular momentum. Note that galaxy interactions 
(among galaxies with relative velocity $V_{rel}$) 
include not only merging but also fly-by events. 
Part of energy released by SNae following star formation is fed back onto the 
galactic gas, thus  returning part of the cooled gas to the hot gas phase

\subsection{Accretion onto SMBHs and AGN emission}

The model also includes a treatment of SMBHs growing at the centre of galaxies
by interaction-triggered inflow of cold gas, following the physical model proposed
by Cavaliere \& Vittorini (2000) and implemented in Menci et al. (2006, 2008). 
The accretion of cold gas is triggered by galaxy encounters (both of fly-by and
of merging kind), which destabilize part of the cold gas available by inducing
loss af angular momentum. The rate of such  interactions is 
\begin{equation}
\tau_r^{-1}=n_T\,\Sigma (r_t,v,V_{rel})\,V_{rel}.
\end{equation}
Here $n_T$ is the number density of galaxies in the same halo,
$V_{rel}$ is their relative velocity, and $\Sigma$ is the cross section for
such encounters which is  given by Saslaw (1985) in terms of the tidal radius
$r_t$ associated to a galaxy with given circular velocity $v$ (see
Menci et al. 2003, 2004).

The fraction of cold gas accreted by the BH in an interaction event is computed
in terms the  variation $\Delta j$ of the specific angular momentum $j\approx
Gm/v_d$ of the gas to read (Menci et al. 2003)
\begin{equation}
f_{acc}\approx 10^{-1}\,
\Big|{\Delta j\over j}\Big|=
10^{-1}\Big\langle {m'\over m}\,{r_d\over b}\,{v_d\over V_{rel}}\Big\rangle\, .
\end{equation}
Here $b$ is the impact parameter, evaluated as the average distance of the
galaxies in the halo. Also, $m'$ is the mass of the  partner galaxy in the
interaction,  and the average runs over the probability of finding such a galaxy
in the same halo where the galaxy with mass $m$ is located.
The values of the quantities involved in the average yield values of $f_{acc}\lesssim 10^{-2}$. 
For minor merging events and for the encounters among galaxies with very unequal mass ratios
$m'\ll m$, dominating the statistics in all hierarchical models of galaxy formation, the 
accreted fraction takes values $10^{-3}\lesssim f_{acc}\lesssim 10^{-2}$. 

The average amount of cold gas accreted during an accretion episode is thus
$\Delta m_{acc}=f_{acc}\,m_c$, and the duration of an accretion episode, i.e.,
the timescale for the QSO or AGN to shine, is assumed to be the crossing time
$\tau=r_d/v_d$ for the destabilized cold gas component.

The  time-averaged bolometric luminosity so produced by a QSO hosted in a given galaxy 
is then provided  by
\begin{equation}
L={\eta\,c^2\Delta m_{acc}\over \tau} ~.
\end{equation}
We adopt an energy-conversion efficiency $\eta= 0.1$ (see Yu \&
Tremaine 2002), and derive the X-ray luminosities $L_X$ in the 2-10
keV band  from the bolometric corrections given in Marconi et al.
(2004). The SMBH mass $m_{BH}$ grows mainly through accretion
episodes as described above, besides  coalescence with other SMBHs
during galaxy merging. As initial condition, we assume small seed
BHs of mass $10^2\,M_{\odot}$ (Madau \& Rees 2001) to be initially
present in all galaxy progenitors; our results are insensitive to
the specific value as long as it is smaller than some
$10^5\,M_{\odot}$.


\subsection{AGN feedback}

Finally, our SAM model includes a detailed treatment of AGN feedback. 
This is assumed to stem from the fast winds with velocity up to
$10^{-1}c$ observed in the central regions of AGNs
(Weymann, Carswell \& Smith 1981; Turnshek et al. 1988; Crenshaw et al. 2003; 
 Chartas et al. 2002; Pounds et al. 2003, 2006; Risaliti et al. 2005); 
these are usually though to originate from the acceleration of disc outflows due to the AGN
radiation field (Proga 2007 and references therein,  Begelman 2004). 
These supersonic outflows compress the gas into a blast wave terminated by
a leading shock front, which  moves outwards with a lower but still
supersonic speed and sweeps out the surrounding medium. Eventually,
this medium is expelled from the galaxy. 

Quantitatively, the energy injected into the galactic gas 
in such inner regions is taken to be proportional to the energy radiated by the 
AGN, $\Delta E = \epsilon_{AGN}\,\eta\,c^2\,\Delta m_{acc}$.  
The value of the energy feedback efficiency for coupling with
the surrounding gas is taken  as $\epsilon_{AGN}=5\, 10^{-2}$ (see Menci et
al. 2008).




\begin{figure*}
\begin{center}
\scalebox{0.30}[0.30]{\rotatebox{0}{\includegraphics{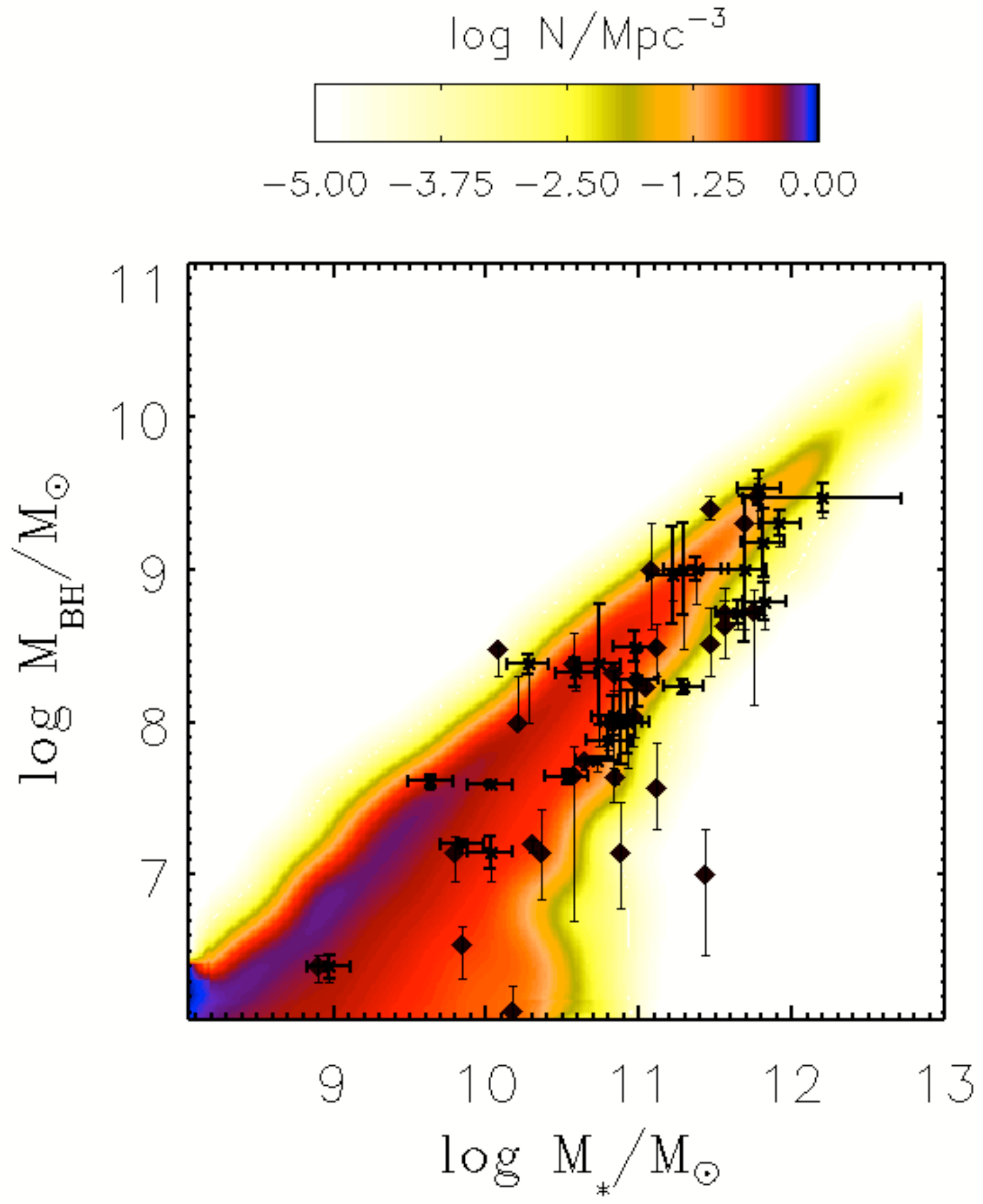}}}
\scalebox{0.30}[0.30]{\rotatebox{0}{\includegraphics{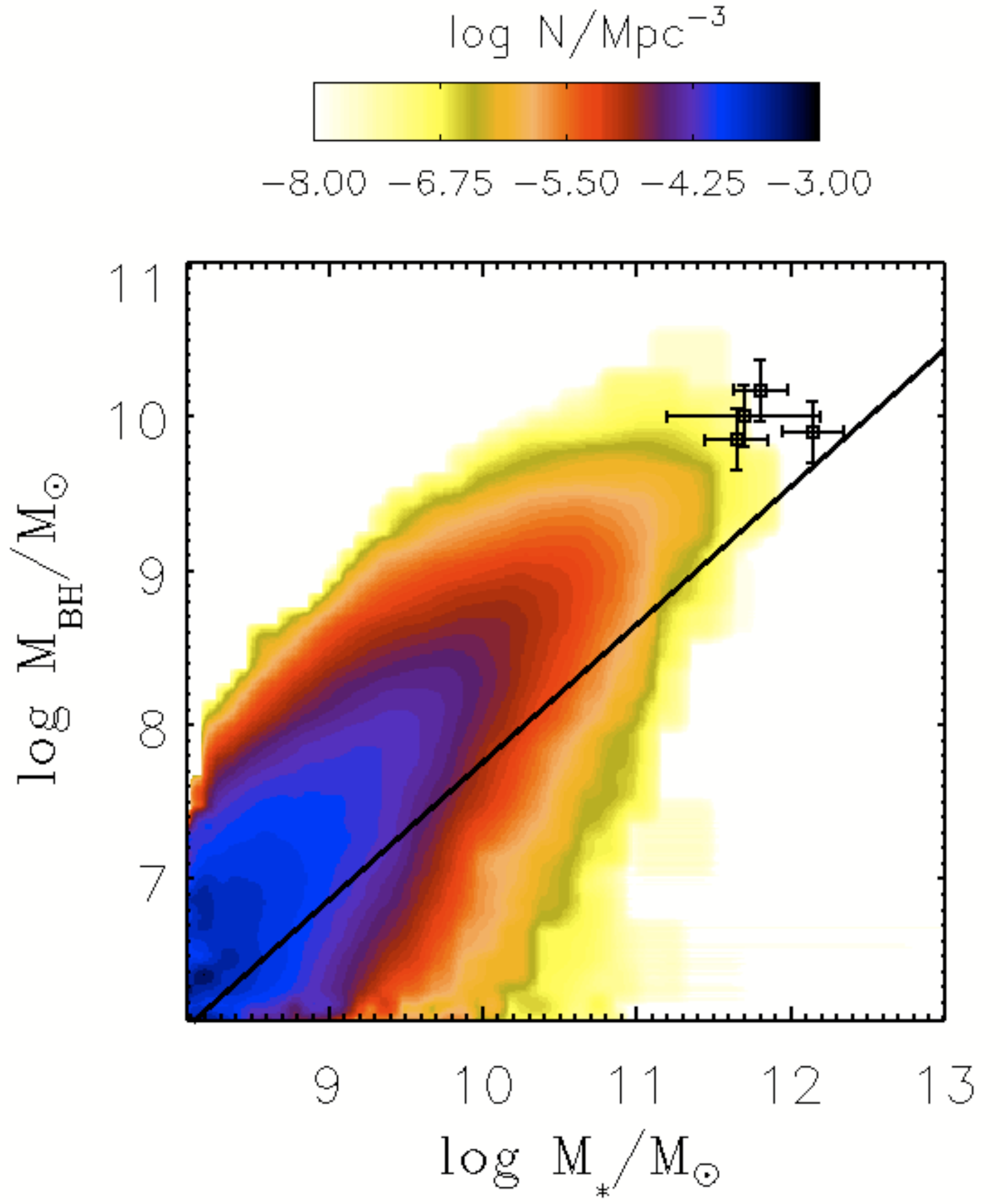}}}
\scalebox{0.32}[0.32]{\rotatebox{0}{\includegraphics{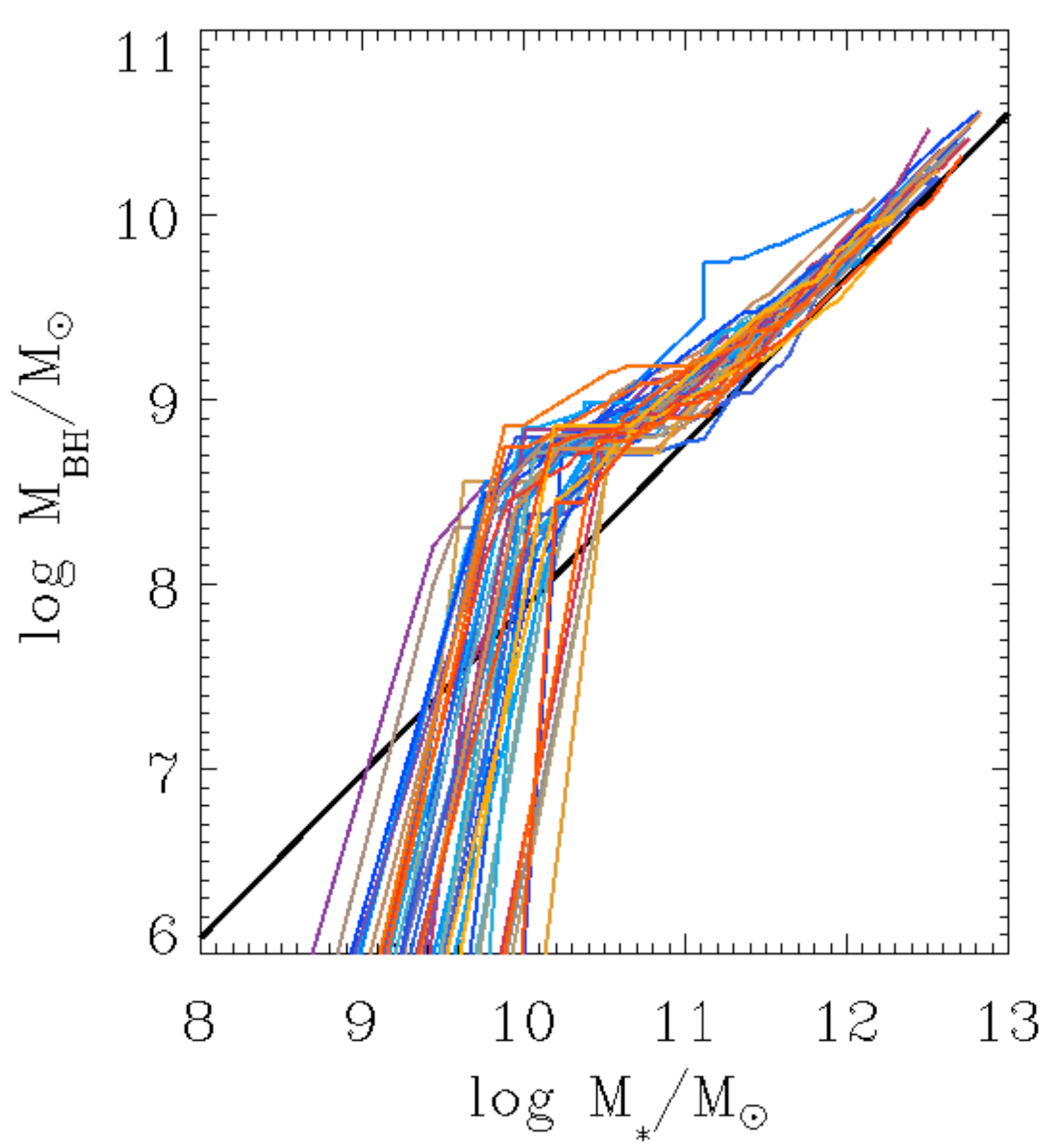}}}
\end{center}
\caption {
Left Panel: The local predicted $M_{BH}-M_*$ relation is compared with data by H\"aring and Rix (2004, diamonds), and 
Marconi \& Hunt (2003, squares, here $M_*$ is derived using the best-fitting virial relation of Cappellari et al. 2006); the colour code represents the density (per $Mpc^3$) of BHs in a given $M_{BH}-M_*$ bin, as indicated by the upper colour bar. 
The contour plot allows to show how the predicted behaviour of the scatter
(growing with decreasing stellar mass) is consistent with the observed scatter of the data points.\newline
Central Panel: The predicted $M_{BH}-M_*$ relation at $z=4$ (colour coded as above) is compared with data 
corresponding to observations of broad-line AGNs and host galaxies in the range $4\leq z\leq 6$; details on the data and corresponding references are provided in the Appendix A. We also show as a solid line 
the predicted median local value, corresponding to the median local value of $M_{BH}$ for model galaxies with given $M_*$.
\newline
Right Panel: We  show some of the paths in the $M_{BH}(t)-M_*(t)$ plane followed, during their evolution, by BHs (and by their host galaxies) reaching a final  mass of $M_{BH}(z=0)\geq 10^{10}\,M_{\odot}$.}
\end{figure*}

\section{Results: Overview of SMBH growth}
As a first step in the study of the relative growth of BHs and host galaxies, we show in fig. 1 the local (left panel) and the 
high-redshift ($z=4$, central panel) $M_{BH}-M_*$ relation that we obtain from our model. 

The predicted local relation 
is consistent with observations, and the data lie on the predicted confidence region represented by the contour plot, although the predicted distribution shows a small 
offset from the observed best-fitting relations derived by Marconi \& Hunt (2003) and H\"aring \& Rix (2004). Although, in principle, the model parameters (like the AGN feedback efficiency $\epsilon_{AGN}$  or the normalization of the star formation efficiency) could be tuned to optimize the fitting to the observed local $M_{BH}-M_*$ relation, this would also affect the predicted properties of the galaxy population (e.g., the evolution of the luminosity function, the colour distributions, the local Tully-Fisher relation) at both low and high redshifts. Since the model is intended to provide a unified description of AGNs and
galaxy evolution, we chose to adopt the same fiducial model adopted in our
previous works since it provides a good match to a wide set of galactic
properties (see Menci et al. 2006, 2008). 
Instead, a robust feature of the $M_{BH}-M_*$ is constituted by the behaviour of the scatter, which is almost independent of the model parameters and even on the specific mechanism assumed to trigger  the AGN feeding. In fact, its decrease with increasing $M_{BH}$ is typical of hierarchical scenarios, and it is due to the early assembly of progenitors into a unique main progenitor which characterizes the merging histories of massive objects formed in biased, high density environments (see Menci et al. 2008). 

Note that  the $M_{BH}-M_*$ relation is predicted to evolve with $z$ as shown by the central panel of fig. 1, with the massive end of the distribution reaching the region corresponding to the  
observations of luminous, broad-line AGNs at $z\geq 4$ (see Appendix A for details on the data). 
A typical prediction of our interaction-driven 
model is that the evolution of the $M_{BH}-M_*$ relation increases for increasing stellar or BH mass; indeed, 
while low-mass BHs in model galaxies at high redshifts are characterized by $M_{BH}/M_*$ ratios close
to their local values (at least for the majority of them), for the  most massive simulated BHs the $M_{BH}/M_*$  is substantially larger than the model local values.
 We can quantify the evolution of the $M_{BH}/M_*$  in our model 
through the quantity $\Gamma$ defined in Sect. 1 (eq. 1); since our 
model predicts a whole distribution of local $M_{BH}$ for each stellar mass $M_*$ 
(not just a simple power-law, as is generally adopted to fit the local data), we compute $\Gamma$ as the 
deviation of the $M_{BH}/M_*$ ratio of each model black hole from its own local value.
Then for galactic hosts with stellar mass $M_*\approx
10^{11}\,M_{\odot}$ the BH masses at $z\geq 4$ are typically $M_{BH}\approx
10^{9.6}\,M_{\odot}$ (see central panel of fig. 1), more
massive than the predicted local value  by a factor $\Gamma$ $\approx$ 5.
These objects constitute the progenitors of the local extremely massive BHs, and their growth with time is represented by the paths shown in the right panel of fig. 1.  Note how such paths are characterized by an assembly of BH masses extremely rapid in the early phases, which is faster than the stellar mass growth since  they approach their final position in the $M_{BH}-M_*$ plane passing above the local $M_{BH}-M_*$ relation. 

The physical origin of the above behaviour can be understood as follows: 
such high-redshift massive BHs (the counterpart of those observed by Walter et
al. 2004; Maiolino et al. 2007; Riechers et al. 2008)  are formed from
galaxies collapsed in biased, high-density regions of the primordial density field where the collapse and
growth of galactic hosts is accelerated.
The  star formation and BH accretion histories of such objects are shown in fig. 2, and can be schematically divided in two phases. 

At high redshifts $z\gtrsim 3$, the rapid interactions characterizing the dense environment of such galaxies are effective in rapidly destabilizing the galactic gas and in feeding the central BHs. 
In fact, in such environments and cosmic times, both the interaction rate in eq. (2) and the fraction of destabilized gas (eq. 3) 
are large; the first, due to the large densities, and the second due to the large ratio $m'/m\approx 1$ (i.e., a mass $m'$ of the merging partner 
close to the mass $m$ of the main progenitor) characteristic of this early phase
($z\gtrsim 3$) when galaxy interactions mainly involve partners with comparable mass. The frequent starbursts in star formation and the corresponding episodes of BH mass accretion at this early cosmic times are shown in fig. 2. 

At lower $z$, the decline of the interaction rate and of 
the destabilized fraction $f_{acc}$ suppresses the growth of BHs which in our model is only due to galaxy interactions (see the drop in 
the BH accretion episodes in the bottom panel of fig. 2), while 
quiescent star formation still proceeds, as is shown by the smooth component in the star formation histories in the top panel of fig. 2.
The latter, quiescent component of star formation continues to build up stellar mass at $z\lesssim 2$ (though at a milder rate), 
thus lowering the $M_{BH}/M_*$ ratio. 
The overall result is that, when the $M_{BH}/M_*$ ratio is normalized to the
final local value, the excess $\Gamma(z)$ increases with redshift. 

\begin{figure}
\begin{center}
\scalebox{0.40}[0.40]{{\includegraphics{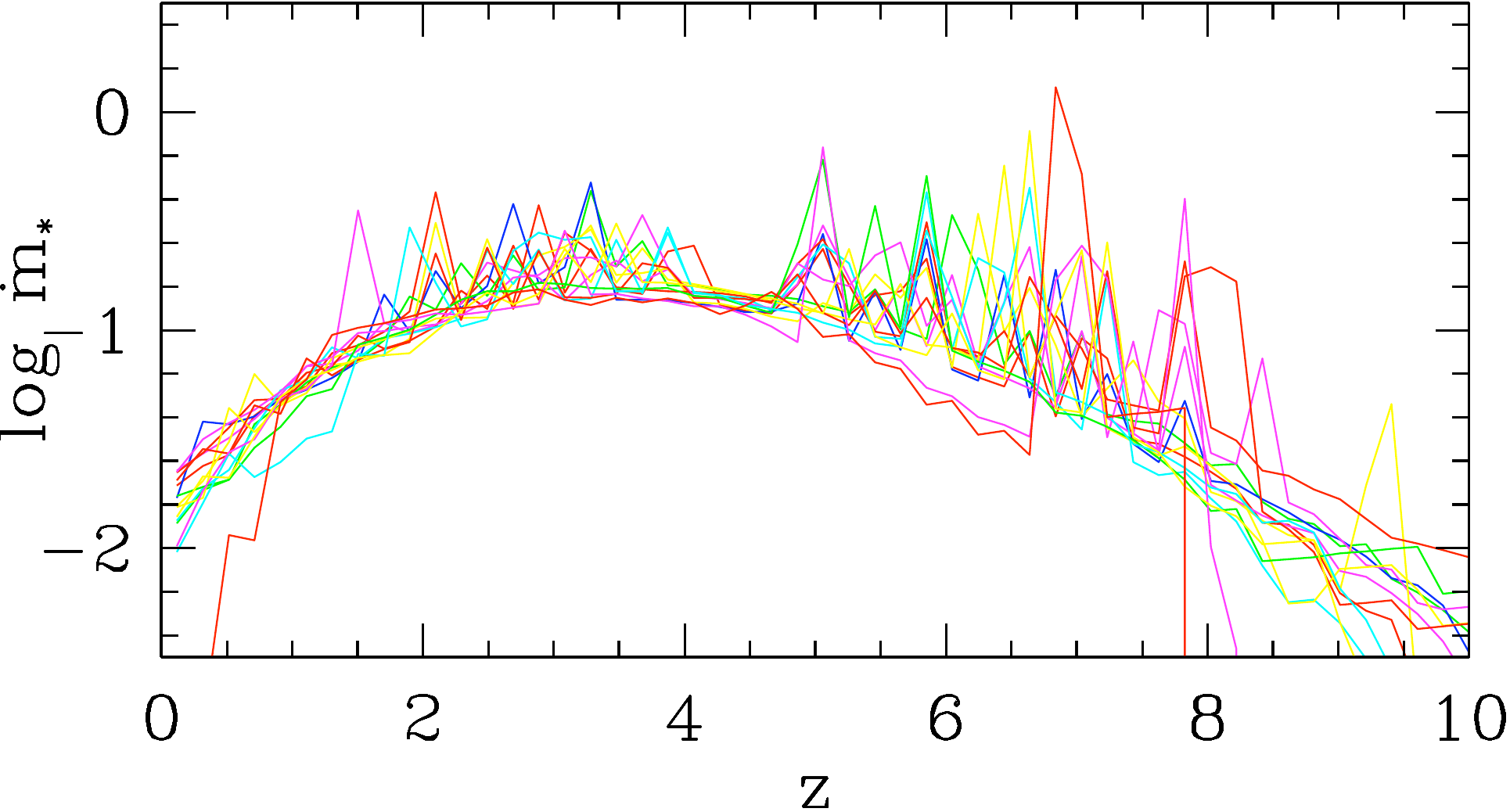}}}
\scalebox{0.40}[0.40]{{\includegraphics{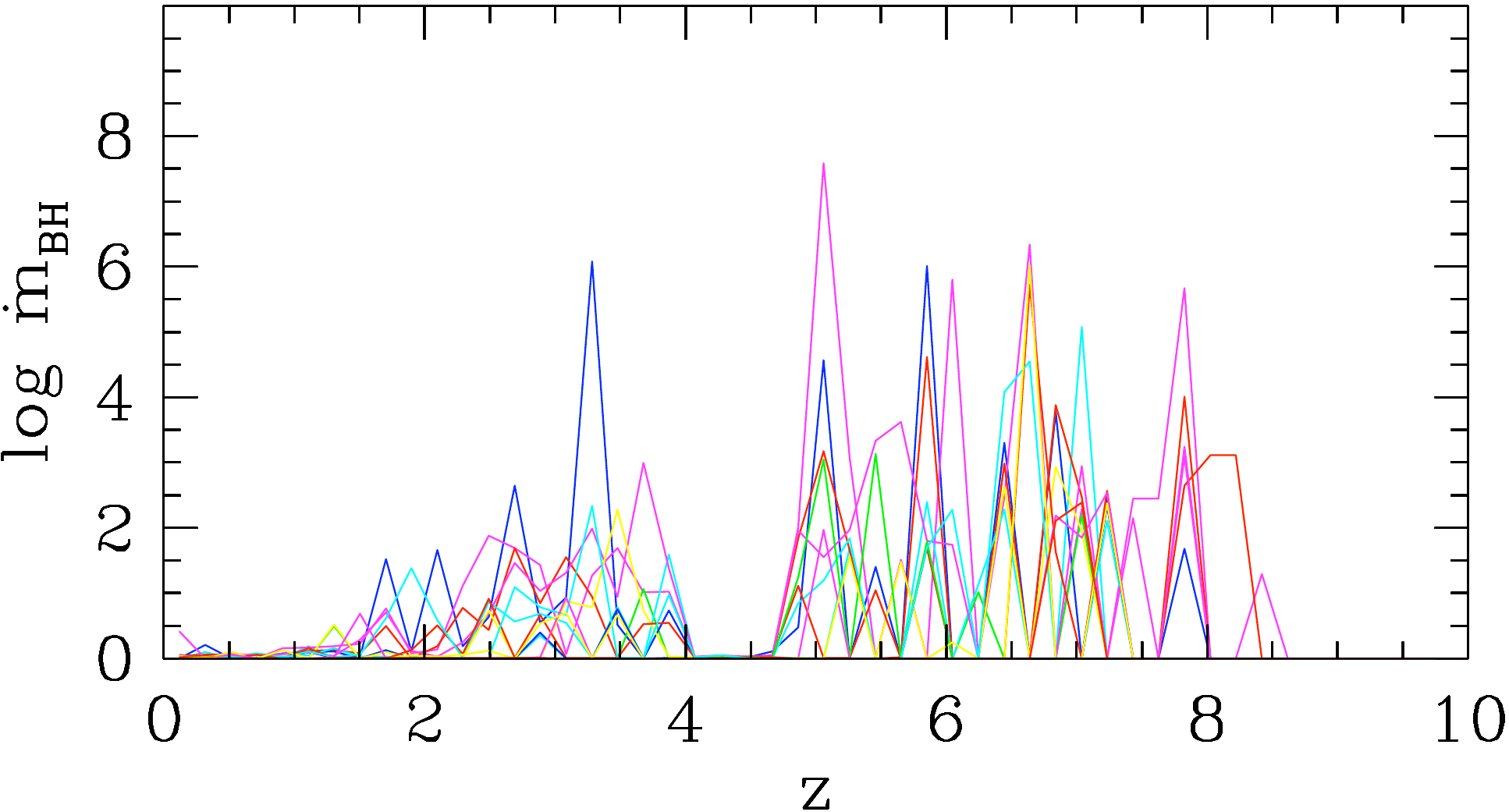}}}
\end{center}
\caption{
Top Panel: We show the star formation histories for a subset of model galaxies. All histories have been normalized to unity when integrated over cosmic time, to make easier the comparison, such a normalization defines the units of the y-axis. \newline
Bottom Panel: The BH mass accretion histories for the same model galaxies, normalized as above. 
The chosen subset of galaxies in our Monte Carlo simulation is the same adopted to illustrate the paths in the $M_{BH}-M_*$ plane 
in fig. 1. }
\end{figure}

Note that the above result is the outcome of two physical processes: 1) the approximatively two-phase growth of cosmic structures 
(with major merging  at high redshifts and accretion of small clumps at low $z$) 
characteristic of hierarchical scenarios, as noted by several authors (e.g., Zhao, Jing \& B\"orner  2003; Diemand, Kuhlen \&  Madau 2007; Hoﬀman et al. 2007; Ascasibar \& Gottloeber 2008); 2) the interaction-driven scenario for the triggering of AGN activity. 

In the above picture, we expect the paths with $\Gamma(z)\geq 1$ to dominate the growth histories of massive objects, formed in biased regions of the density field where high-redshift interactions are extremely effective, while we expect the high-redshift values of $\Gamma$ to progressively lower when galaxies formed in less dense environment (and hence with a lower mass on average at any given $z$) are considered, due to the progressively lower efficiency of  interactions in triggering BH accretion at high-$z$. 
Indeed, the dependence of the above effect on BH mass and on redshift constitutes a typical signature of the above 
interaction-driven model for the growth of BHs in galaxies.



\section{Results: The Evolution of $M_{BH}-M_*$ for Different AGN Populations}

\subsection{The Early BH Growth and High-redshift QSO}

An efficient way to test the above scenario is to investigate how the paths followed by BHs and by their host galaxies to reach the local  $M_{BH}-M_*$ 
relation depend on their properties (mass, gas fraction, star formation rate), and to compare such specific predictions with the corresponding available observations  for which we will use  the mean local value of the $M_{BH}/M_*$  ratio obtained from the H\"aring \& Rix (2004) sample  to calculate $\Gamma$.
We shall devote to this the next subsections.

\begin{figure}
\begin{center}
\scalebox{0.35}[0.35]{{\includegraphics{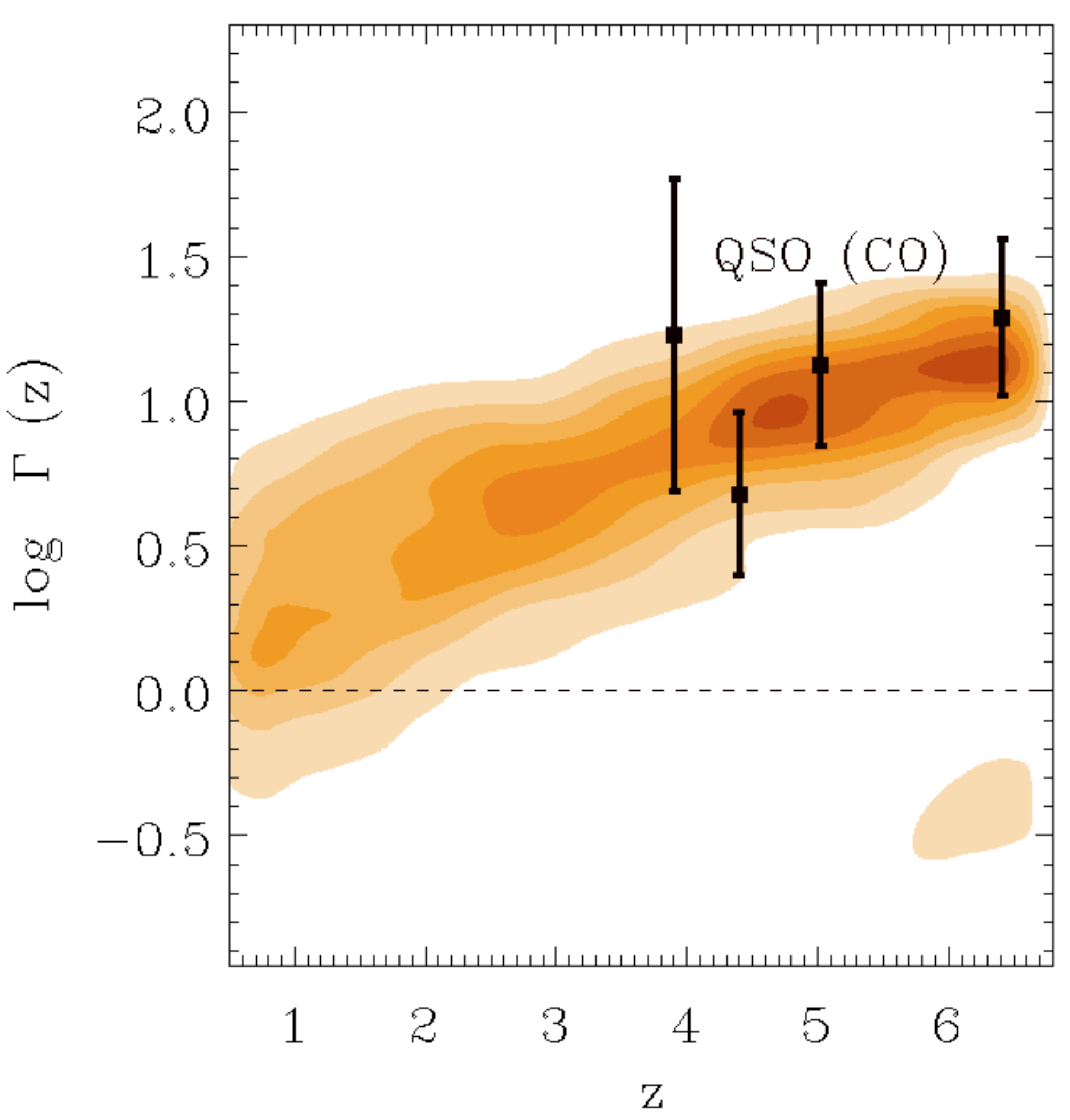}}}
\end{center}
\caption{
The predicted evolution of the BH to stellar mass ratio $\Gamma$ (normalized to the local value, see eq. 1), for 
the evolution of BHs with masses $M_{BH}\geq 10^9\,M_{\odot}$ at $z\geq 4$. The 6 filled contours correspond to equally 
spaced values of the fraction of objects with a given value of $\Gamma(z)$ at the considered redshift: from 0.01 for the lightest filled region  to 0.1 for the darkest. For reference we draw as a dashed line the local value $\Gamma(0)=1$. The data points represent the excess $\Gamma$ derived from the observed BH masses of 
the high-redshift broad-line AGN following the procedure described in  Appendix A. }
\end{figure}

\begin{figure*}
\begin{center}
\scalebox{0.32}[0.32]{{\includegraphics{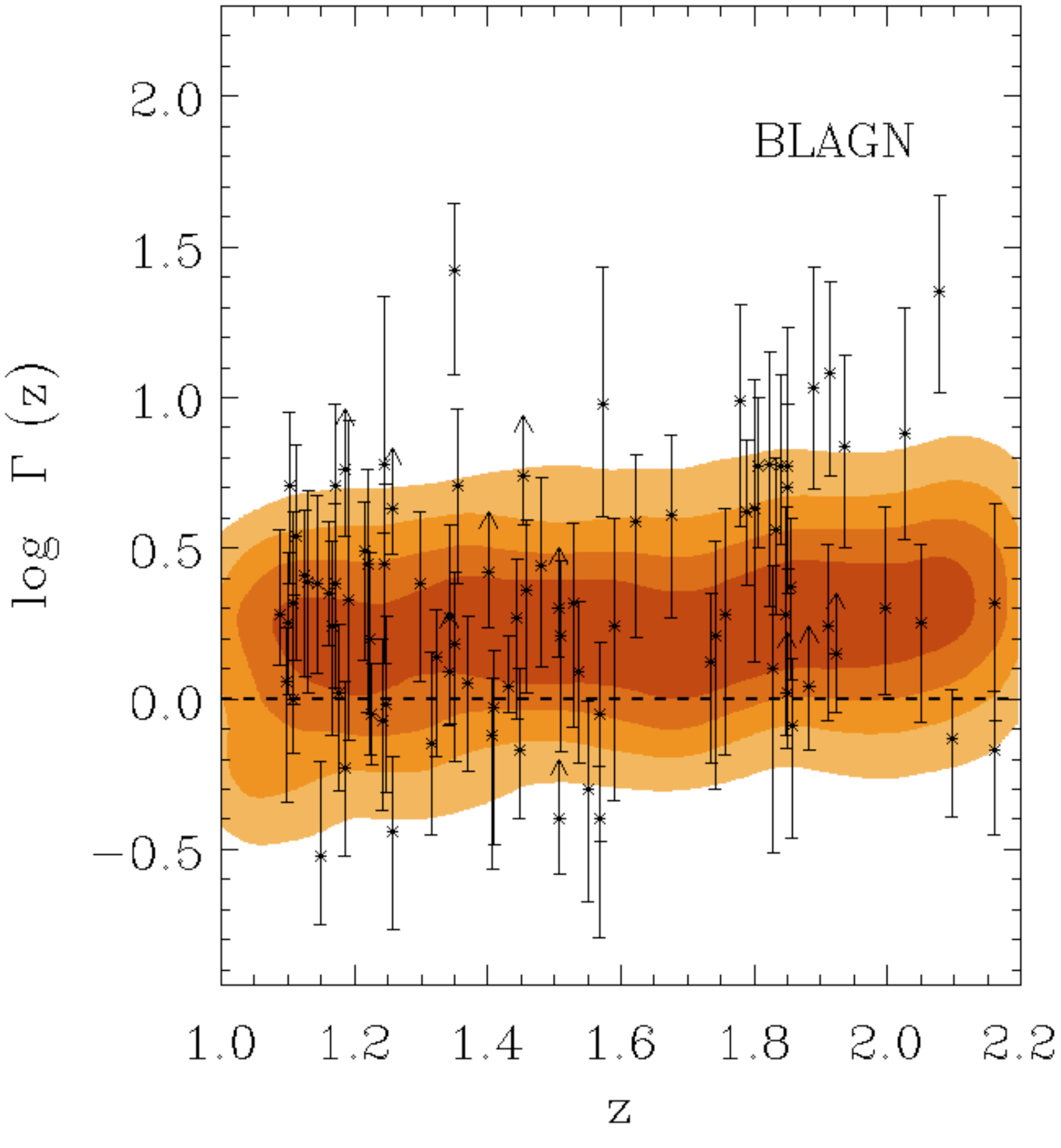}}}
\scalebox{0.32}[0.32]{{\includegraphics{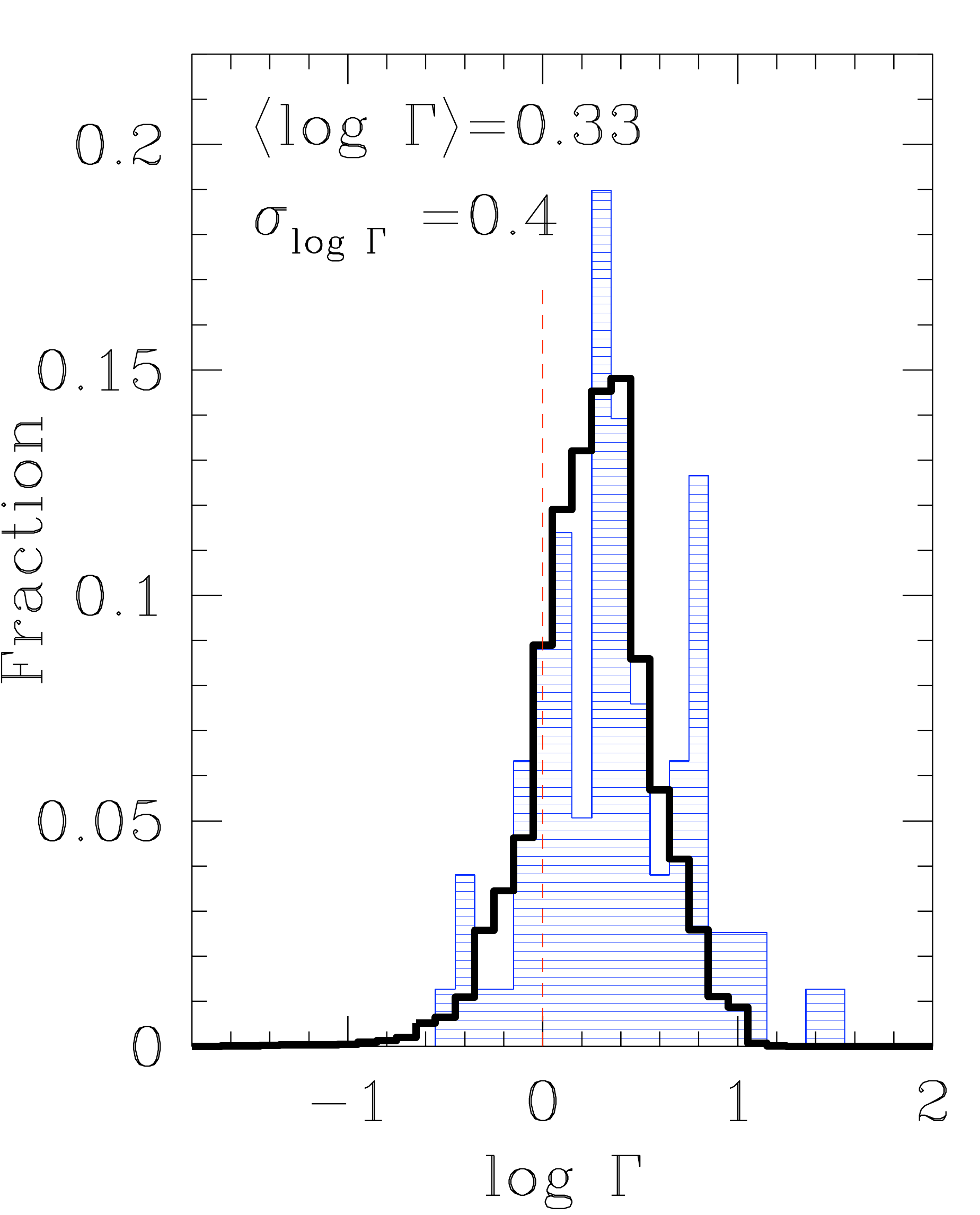}}}
\end{center}
\caption{
Left Panel: The predicted evolution of the BH to stellar mass ratio $\Gamma$ (normalized to the local value, see eq. 1), for AGNs selected as to have a bolometric luminosity $L\geq 10^{44.5}$ erg/s at $1\leq z\leq 2$.  
The 4 filled contours correspond to equally 
spaced values of the fraction of objects with a given value of $\Gamma(z)$ at the considered redshift: from 0.01 for the lightest filled region  to 0.1 for the darkest. 
The data points represent the excess $\Gamma$ derived from the BH masses of the COSMOS broad-line AGN measured by Merloni et al. (2009).  Note that the bulk of the BH building up occurred ad higher redshift, when most of the 
galaxy interactions triggering the BH accretion take place: the slow evolution $\Gamma(z)$ for $1\leq z\leq 2$ is 
driven by the gradual and much milder growth of the stellar mass. 
\newline
Right Panel: The distribution of $log\,\Gamma$ for all galaxies in the redshift range $1\leq z\leq 2$ for model galaxies (solid line) is compared with the corresponding observed distribution (Merloni et al. 2009, filled  histogram): the latter has been obtained including the lower limits shown in the left panel. The average value and the dispersion for $log\,\Gamma$ shown on the top refer to the observed distribution, these values are nearly insensitive to the exclusion of the lower limits.}
\end{figure*}

To explore in detail the growth of BHs in galaxies with different properties, we start with the most extreme objects, namely, the 
massive BH already in place at high redshifts. 

The time evolution of the $M_{BH}/M_*$ ratio of such extreme objects is shown
in fig. 3 in terms of the excess $\Gamma(z)$ over their local value  for model
BHs with masses  $M_{BH}\geq 10^9\,M_{\odot}$ at $z\geq 4$. For such objects
$\Gamma(z)\geq 1$ holds at any time between $z=6$ and the present, with
$\Gamma\approx 5$ holding at at $z=5-6$. These $\Gamma$ values are consistent
with the observations of luminous QSO at 3.6$<z<$6.4 (see Appendix A).


\subsection{The BH growth at Intermediate-redshift and BL AGNs}


Now, we focus on BL AGNs at intermediate redshift $1\leq z\leq 2$ and AGN bolometric luminosity $L\geq 10^{44.5}$ erg/s, for which the model results can be compared with detailed  existing observations. Such objects are expected to form in less biased regions 
of the primordial density field compared to the class of high-redshift QSO
discussed in the previous section; in fact, they have a comparable AGN luminosity but are found at a lower redshifts, and thus are expected to originate from merging histories characterized by a 
lower rate of high-redshift encounters. Therefore, according to our picture discussed at the end of Sect. 3, we expect such objects to show a lower  excess $\Gamma (z)$. This is indeed the case, as illustrated in fig. 4, where we compare the model predictions with 
observations by Merloni et al. (2009).

The distribution of $\Gamma$ for such intermediate objects is still dominated by $\Gamma\geq 1$ at any refdshift; the model predicts typical values $\Gamma\approx 1.5-2$, smaller than those attained by the high-redshift QSO shown in fig. 3, and consistent with the observational range. 
Besides  providing an average $\Gamma$ close to the observational value, it is interesting to note how the model predicts a {\it scatter} in remarkable agreement with the observed distribution (see right panel in fig. 4). This indeed constitute an extremely important test for models based on the hierarchical scenarios, since the scatter is directly related by the spread in the merging histories of the host galaxies, a 
specific prediction of these models which cannot be tuned though adjustable free parameters. 

Note that  the model predictions shown in fig. 4 refer to {\it all} AGNs in
the  redshift and luminosity ranges specified above, while the data we compare
with include only unobscured objects. However, in our model the obscuration
properties of AGNs depend only on the amount of gas swept by the blast wave originated by the AGN at the time of observation (see Sect. 2.3), i.e., on the detailed hydrodynamic of the interstellar medium during the last, 
short ($\Delta t\lesssim 5\,10^7$ yrs) AGN active phase. Since the $\Gamma$ ratios of model BHs basically depend on the integrated, past history of the simulated BHs and host galaxies, we expect distribution of $\Gamma$ for obscured and unobscured AGNs to be similar. While we cannot perform an exact investigation of this issue (we do not model the nuclear obscuration of AGNs), a first-order estimate of the effect of selecting only 
obscured objects can be performed by considering in our model only AGNs hosted in galaxies with high gas surface densities $\Sigma_{gas}\geq 30\,M_{\odot}$ pc$^{-2}$: even in this case the mean value of $\Gamma$ in the distribution in fig. 4 is only decreased by a factor $\approx 1.2$, leaving the scatter unchanged.


\subsection{The Growth of Black Holes in Gas-Rich Galaxies: Comparison with  Submillimiter Emitting Galaxies} 

In our model, a completely different set of BH accretion histories is that corresponding to AGNs in sub-mm galaxies (SMGs). Their large gas fractions  ($f_{gas}\gtrsim 0.6$ relative to the total baryonic mass) and star formation rate ($\dot m_* =100-1000$ $M_{\odot}$/yr, see Introduction) 
indicate that these galaxies originate from merging histories characterized by less prominent 
high-redshft ($z\gtrsim 4$) starbursts and BH accretion episodes, so that a large fraction of gas is left available at lower redshifts 
$z\approx 2-3$. To compute quantitatively the predicted $M_{BH}$ and $M_*$ for such galaxies and to compare with existing 
observations, we selected from our Monte Carlo simulations galaxies with i) gas fractions $f_{gas}\geq 0.7$, ii)  star formation
 rates $\dot m_*\geq 100$ $M_{\odot}$/yr; iii) AGNs with X-ray luminosities $L_1\leq L\leq L_2$, where $L_1$ 
 and $L_2$ have been chosen as to match the selection criteria adopted for the observations we compare with (Borys et al. 2005; 
 Alexander et al. 2008); the X-ray luminosities (in the band $2-10$ keV) have been computed from the bolometric luminosities in eq. (4) adopting the bolometric correction by Marconi et al. (2004). 
\begin{figure*}
\begin{center}
\vspace{0.cm}
\scalebox{0.35}[0.35]{{\includegraphics{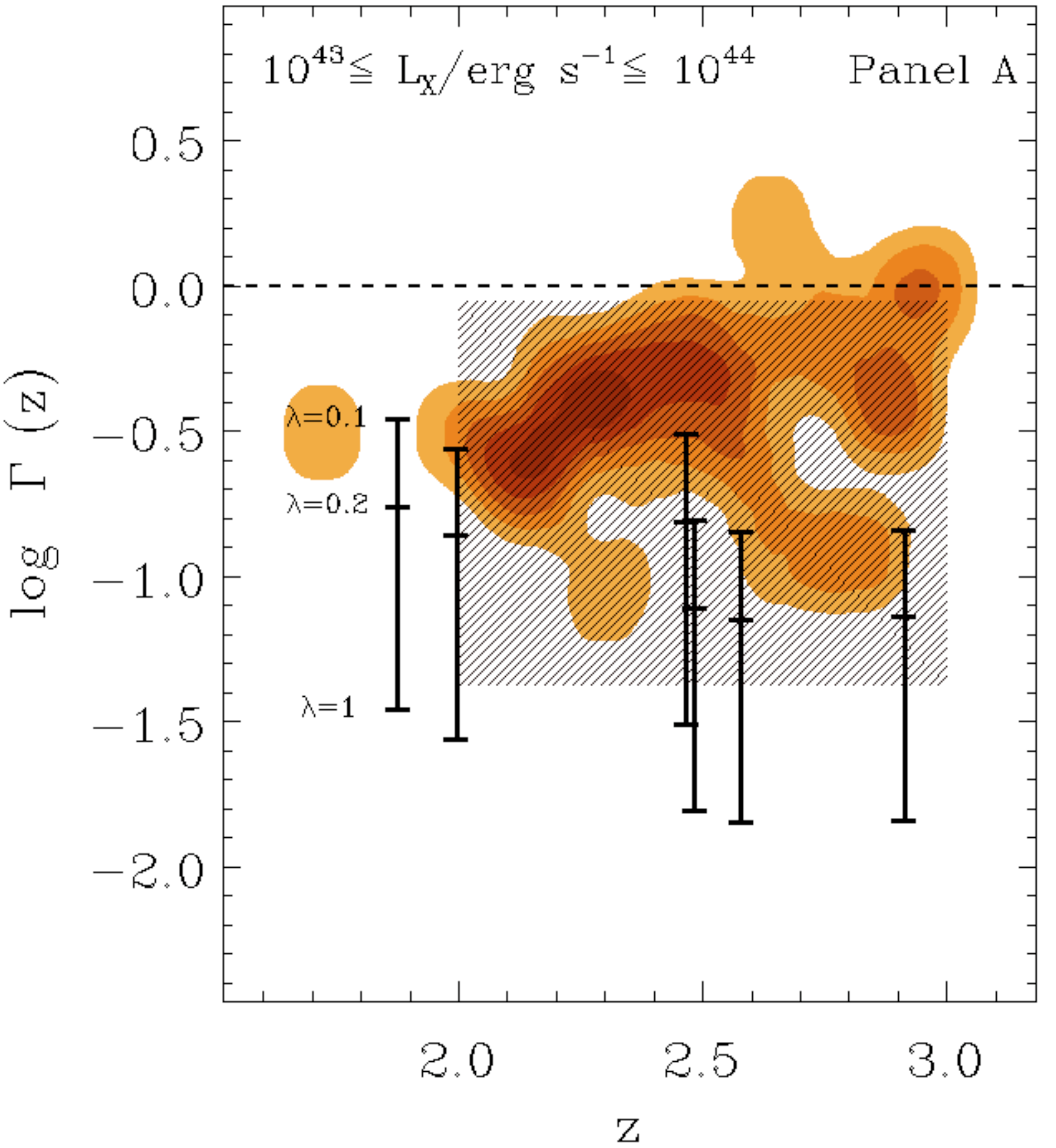}}}
\scalebox{0.35}[0.35]{{\includegraphics{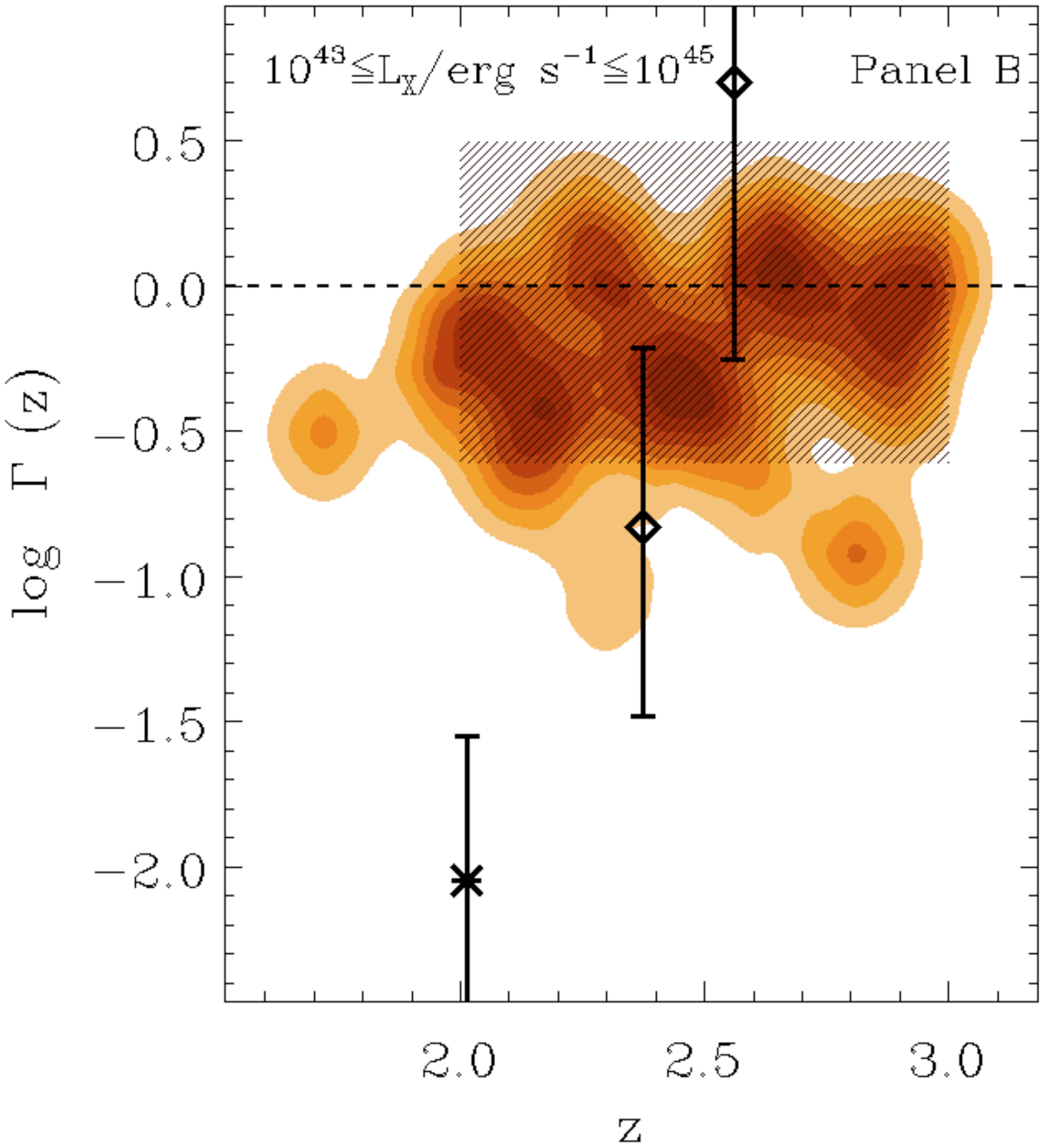}}}
\end{center}
 \caption{The predicted evolution of the BH to stellar mass ratio $\Gamma$ (normalized to the local value, see eq. 1), is shown in Panels A and  B for AGNs in SMG-like galaxies selected from our Monte Carlo simulation according to the criteria i)-iii) (see text). The 5 filled contours correspond to equally 
spaced values of the fraction of objects with a given value of $\Gamma(z)$ at the considered redshift: from 0.01 for the lightest filled region  to 0.1 for the darkest. Panels A and B refer to the different selection in X-ray luminosity of the AGNs shown on the top of the Panels. 
We compare with data for  SMGs with 
X-ray obscured AGNs (Panel A) and  with BL AGNs (Panel B) in the same luminosity range adopted for the selection of model galaxies; 
the data correspond to individual galaxies for which both  BH and stellar
or dynamical masses measurements were available in the literature (see Appendix B). The errorbars in the Panel A show only the uncertainties  
in the data due to the adopted Eddington ratio $\lambda$, while errorbars on Panel B include the observational uncertainties in both the BH and the stellar mass measurements. We also show  as a shaded area the $1-\sigma$ uncertainty region around the value of $\Gamma$ obtained by Alexander et al. (2008) from average BH (adopting $\lambda=0.2$) and stellar masses (obtained from near IR luminosity and CO line widths) taken from various samples (see Alexander et al. 2008 and references therein). Note that the extremely gas-rich galaxies selected as SMGs represent rare evolutionary paths (see text for the computed and the observed number densities) compared to typical galaxies at such redshifts: the statistical fluctuations associated with the low number 
of selected SMGs in our model reflects in the irregular contours and the isolated bins shown in the figure.}
\end{figure*}

The resulting predicted distribution of the $M_{BH}/M_*$ ratio are shown in figs. 5a and 5b for SMG with AGN luminosity in two different ranges 
(see caption) as to compare with the $\Gamma$ estimates of six X-ray obscured
 and three broad line (BL) SMGs for which individual estimates of M$_{BH}$ and M$_*$  were available in the literature
 (see the Appendix B for details on the adopted data).
Although the comparison with observations results are still indicative (due to
the large uncertainties associated with the Eddington ratios and  with the
measurements of both the BH and the stellar mass, see Appendix B), 
the model predicts for bulk of the SMG population values of $\Gamma\leq 1$, due to the slower growth of BHs at high redshifts. This is due to 
the particular merging histories corresponding to our selection criteria; large residual gas fractions at $z\approx 2.5$ imply less  
frequent high-redshift encounters at $z\geq 4$ and less effective bursts and BH accretion episodes at these early epochs. Since 
in our model the high-redshift interactions are the only trigger for early BH growth, this results in lower values of $\Gamma(z)$ at later times ($z=2-3$).  
Such results do not depend on the details of the selection criteria i)-ii) that we adopted to extract 
SMG-like galaxies from our Monte Carlo simulations.  Indeed, we have verified that adopting different thresholds for the gas fractions ($f_{gas}\geq 0.6$) and star formation rates (up to 400 $M_{\odot}$/yr) the fraction of galaxies with $\Gamma$ $<$1 is always dominant.

 It must be noted that the observed SMG at $z$=2 in the right panel of fig. 5 has a $\Gamma$ value much lower than that predicted by the model at the same redshift. In view of the unavoidable uncertainties affecting the present observational data (discussed in Appendix B), it will be important to confirm the reliability and the statistical significance of such a data point as a marker of the unobscured AGNs in SMGs, since  it would constitute a severe test for our model 
which predicts a basically similar behaviour of $\Gamma(z)$ for obscured and unobscured AGNs. 
On the other hand, the
predicted distribution of $\Gamma$ for SMG-like galaxies shows a mild dependence with the AGN luminosity, being larger for luminous AGN. 
Although the uncertainties in present data do not allow yet to make precise
comparisons, this trend constitutes a model prediction which is at least
consistent with estimates of $\Gamma$ based on average values for $M_{BH}$ and
$M_*$ in SMG galaxies (see the shaded area). 


An interesting feature of our interaction-driven model for BH accretion is that 
the SMGs produced by the model 
end up in low-redshift descendants which lie {\it below} the local $M_{BH}-M_*$ relation as shown in fig. 6.
In other words, the low-redshift ( $z\leq 2$) descendants of SMGs are predicted to have BHs with low-intermediate masses 
$M_{BH}=10^8-10^9$ $M_{\odot}$ (see fig. 6), and the BHs in SMG-like galaxies at $2\lesssim z\lesssim 3$ are in an active growing phase, 
as it is shown in fig. 6 by their steeply rising paths in the $M_*-M_{BH}$ plane. Conversely, we computed that a fraction $\approx 10\%$ of local galaxies with BH masses in the range $M_{BH}=10^8-10^9$ $M_{\odot}$ had a progenitor which passed through an SMG-phase at some redshift 
in the range $2\leq z\leq 3$.

Such a picture is in agreement with the independent findings of Alexander et al. (2008), 
based on the larger number density of SMGs compared to that of local galaxies hosting BHs with masses exceeding $10^9$ 
$M_{\odot}$. To check for the full consistency of our results and of our interpretation with the observations of SMGs, we have compared the number densities of the model SMG galaxies with the number densities $\rho_{SMG}\simeq$ 2.5$\times$10$^{-5}$Mpc$^{-3}$ measured by Swinbank et al. (2006) based on the redshift distribution obtained by Chapman et al. (2005) for $z$=1-3.5 SMGs with S$_{850\mu m}>$  5mJy. To perform such a comparison we first converted the above flux limit in terms of star formation rate (using the relations in Swinbank et al. 2008), since our semi-analytic model does not include the sub-mm emission from dust; for the model SMG galaxies above the resulting threshold in star formation rate 
the predicted number density in the same redshift interval is $\rho_{SMG}=4.4\times$10$^{-5}$Mpc$^{-3}$ when SMGs with $f_{gas}
\geq 0.6$ are considered, and $\rho_{SMG}=1.9\times$10$^{-5}$Mpc$^{-3}$ for SMGs with 
$f_{gas}\geq 0.7$, fully consistent with the observational estimates. 

\begin{figure}
\begin{center}
\scalebox{0.35}[0.35]{{\includegraphics{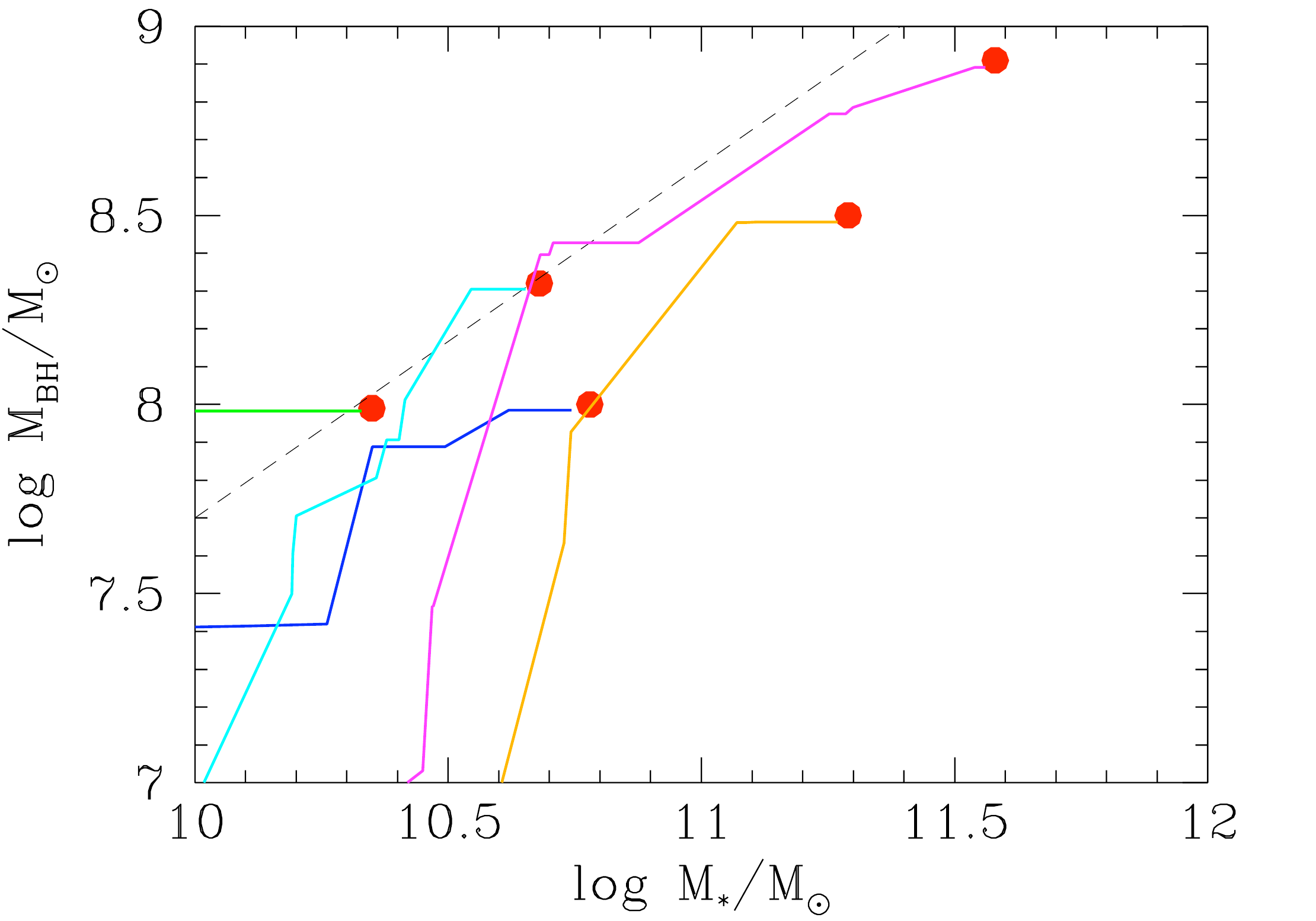}}}
\end{center}
\caption{Typical growth paths for BH and stellar masses for our model SMG-like galaxies. The dots represent the final location (at $z=0$) of the descendents of SMG-like galaxies selected according to our criteria i)-iii) (see text) in the redshift range $2\leq z\leq 3$. The dashed line represent the median value of the local $M_{BH}-M_*$ relation for all model galaxies at $z=0$. }
\end{figure}

\section{Discussion}
\subsection{The growth of BHs for different AGN populations}
The results presented in the previous sections lead to the following global picture of our cosmological, interaction-driven model for the BH accretion. 
The relative growth speed of SMBHs compared to the stellar mass is
characterized by a large spread in the galaxy population, with an overall
trend for a faster BH growth compared to galaxies at high redshifts $z\gtrsim3$
; this is basically due to the fact that BH grow only when interactions are effective, i.e., at high redshifts $z\gtrsim 3$, while star formation proceeds not only through impulsive bursts (at high redshifts) but also through quiescent star formation which continues to build up stellar mass at $z\lesssim 2$, though at a lower rate (see fig. 2a). 
In terms of the ratio $\Gamma(z)$ such a global behaviour is represented by fig. 7, where the distribution of $\Gamma$ is shown by the coloured contours as a function of redshift for the entire (i.e., for all galaxies containing BH with masses $M\geq 10^5\,M_{\odot}$).  A striking feature of such a distribution is the modest increase of $\Gamma(z)$ for $z\lesssim 3$ as opposed to its rapid upturn at higher redshifts $z\gtrsim 4$; such a mild increase of the 
global $\Gamma$ at low-intermediate redshifts is in agreement with that derived by Merloni et al. (2004) and Hopkins et al. (2006) by comparing the observed integrated BH and stellar mass densities at such redshifts. 

\begin{figure*}
\begin{center}
\scalebox{0.38}[0.38]{{\includegraphics{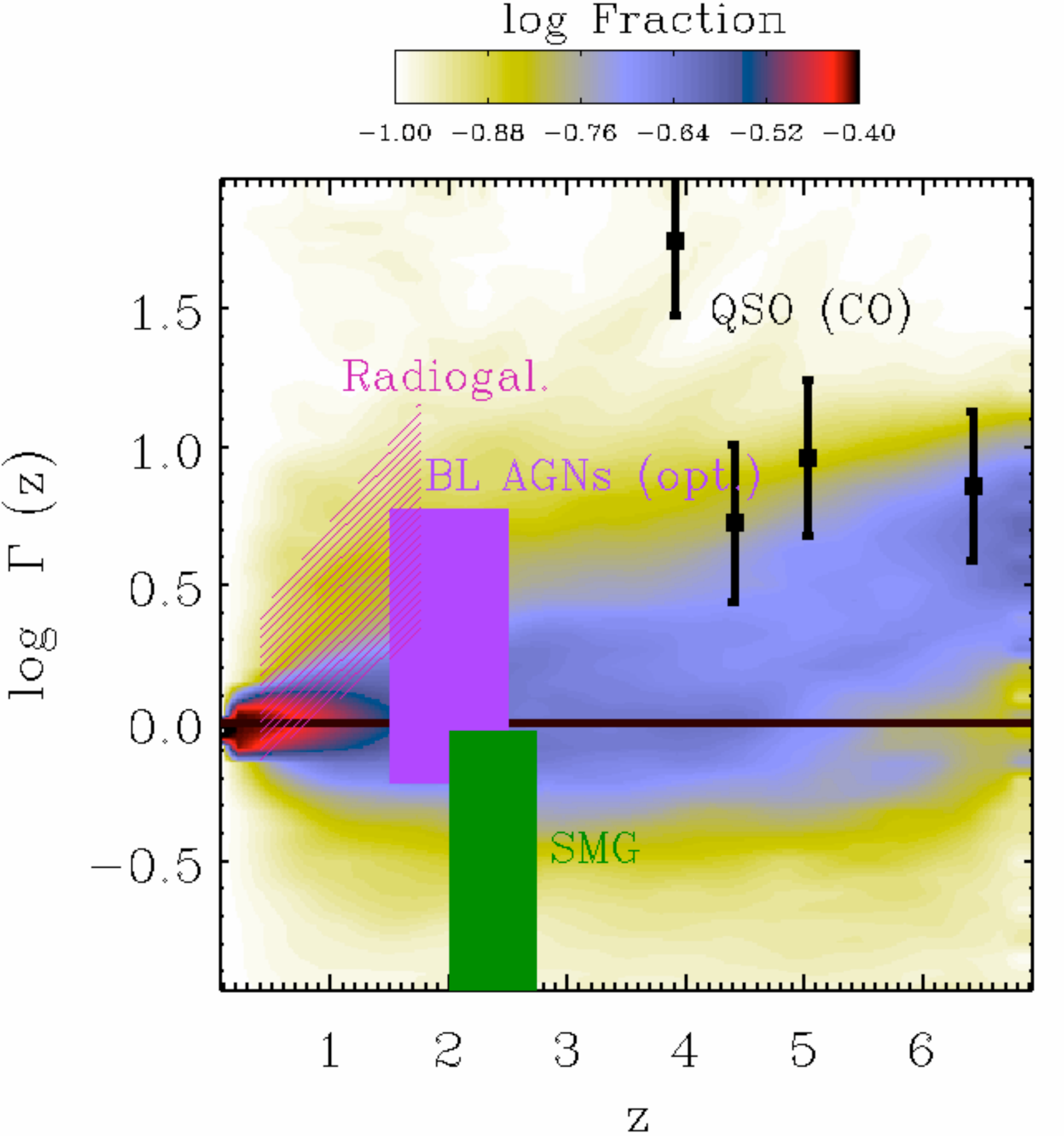}}}
\end{center}
\caption{The predicted evolution of the BH to stellar mass ratio $\Gamma$ for the whole galaxy population in our Monte Carlo simulation; the colour code represents, at any redshift $z$, the fraction of objects with excess $\Gamma(z)$. 
The average values of $\Gamma$ derived from observations of SMGs (Alexander et al. 2008), of BL AGNs (Merloni et al. 2009) and high-redshift QSOs (see Maiolino et al. 2007 and Appendix A) are shown as shaded areas covering their observed redshift range and observational uncertainties. } 
\end{figure*}
Note that the overall predicted trend for larger values of $\Gamma$ with increasing redshift is associated to an increasing the spread in the  distribution.  This constitutes a quantitative prediction for the  $z$-dependence of the Lauer et al. (2007) bias. 
In this respect, the $\Gamma$  distributions shown  in fig. 3, 4,  and 5 can be interpreted as a quantitative evaluation of the incidence of such a bias  in differently selected samples at different redshifts. 
In fact, fig. 7 shows that the different observations we compared with in the previous Sections are not fully representative of the global behaviour since they refer to specific populations selected according to different criteria which sample only a portion (in the case of SMG actually a minor fraction) of the whole galaxy population. 

\subsection{Model predictions and downsizing}
A major result of our previous sections is that the evolution of $\Gamma (z)$ is a {\it strong} function of the BH mass.
Such a strong mass dependence of $\Gamma (z)$ is illustrated by fig. 8 (left panel), where we show with the contours the \textit{average} values $\Gamma(z) $  as a function of the final (z=0) BH mass (represented in the y-axis) and of redshift; the average runs over all paths leading to the final BH mass  (and hence over all the possible main progenitors at a given redshift). The figure illustrates that indeed for massive objects $M_{BH}\geq 10^9\,M_{\odot}$ at high redshifts $z\gtrsim 4$ large values of $\Gamma>3$ are expected, since such BH form in the most biased regions of the density field where high-redshift interactions were favoured; values of $\Gamma\gtrsim 1.5-2$ are also natural for BH at intermediate redshifts $1\leq z\leq 2$, while the low values of $\Gamma$ observed in SMG at $2\leq z\leq 3$ are not representative of the whole AGN population at such redhifts (the average $\Gamma$ represented in fig. 8 is $\approx 1.5$), but instead only apply to a population  originated by peculiar star formation histories (see section 4.3). 
  
An immediate implication of the above is that {\it massive local galaxies} and their BHs have formed preferentially through paths (in the $M_*-M_{BH}$ plane) passing {\it above} the local  $M_*-M_{BH}$ relation; this is illustrated in fig. 8 (right panel)  where we show the fraction of paths characterized by a given $\Gamma$  (in two redshift bins) leading to local galaxies with mass $M_*\geq 10^{12}\,M_{\odot}$. The paths with $\Gamma>1$ dominate the statistics not only at high-redshift $z=4-5$ but also -though to a lower extent- at $z=1-3$.  As a consequence, in the merging-driven scenario SMGs do not represent typical paths leading to local massive galaxies, but rather correspond to peculiar paths in the tail of the distribution, namely those selected as to lead to gas-rich galaxies at $z=1-3$; this interpretation is supported by the consistency of the predicted number density of such peculiar paths with the observed number densities of SMGs, shown and discussed at the end of sect. 4.3.
  
  \begin{figure*}
\begin{center}
\scalebox{0.33}[0.33]{{\includegraphics{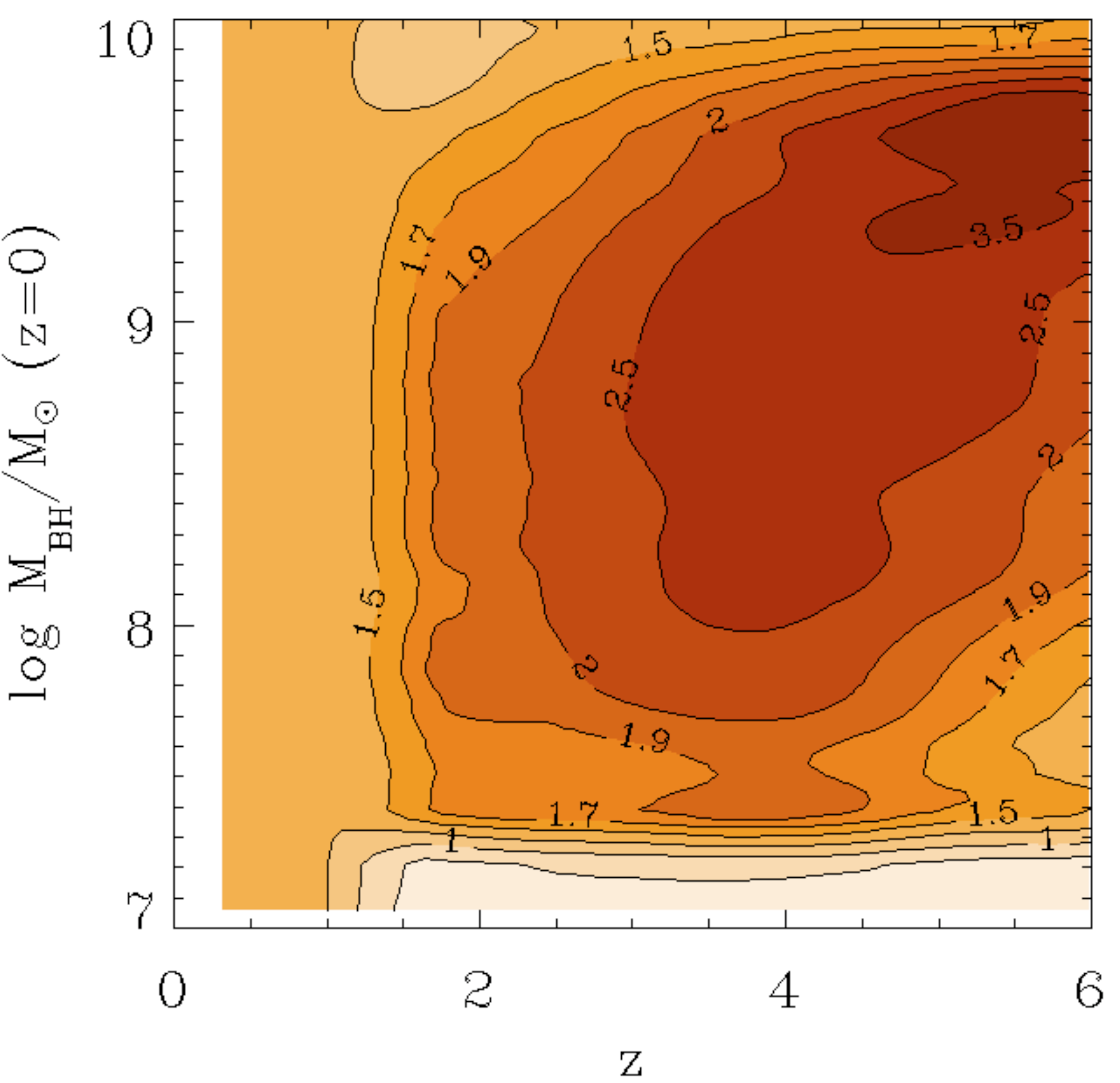}}}
\scalebox{0.36}[0.36]{{\includegraphics{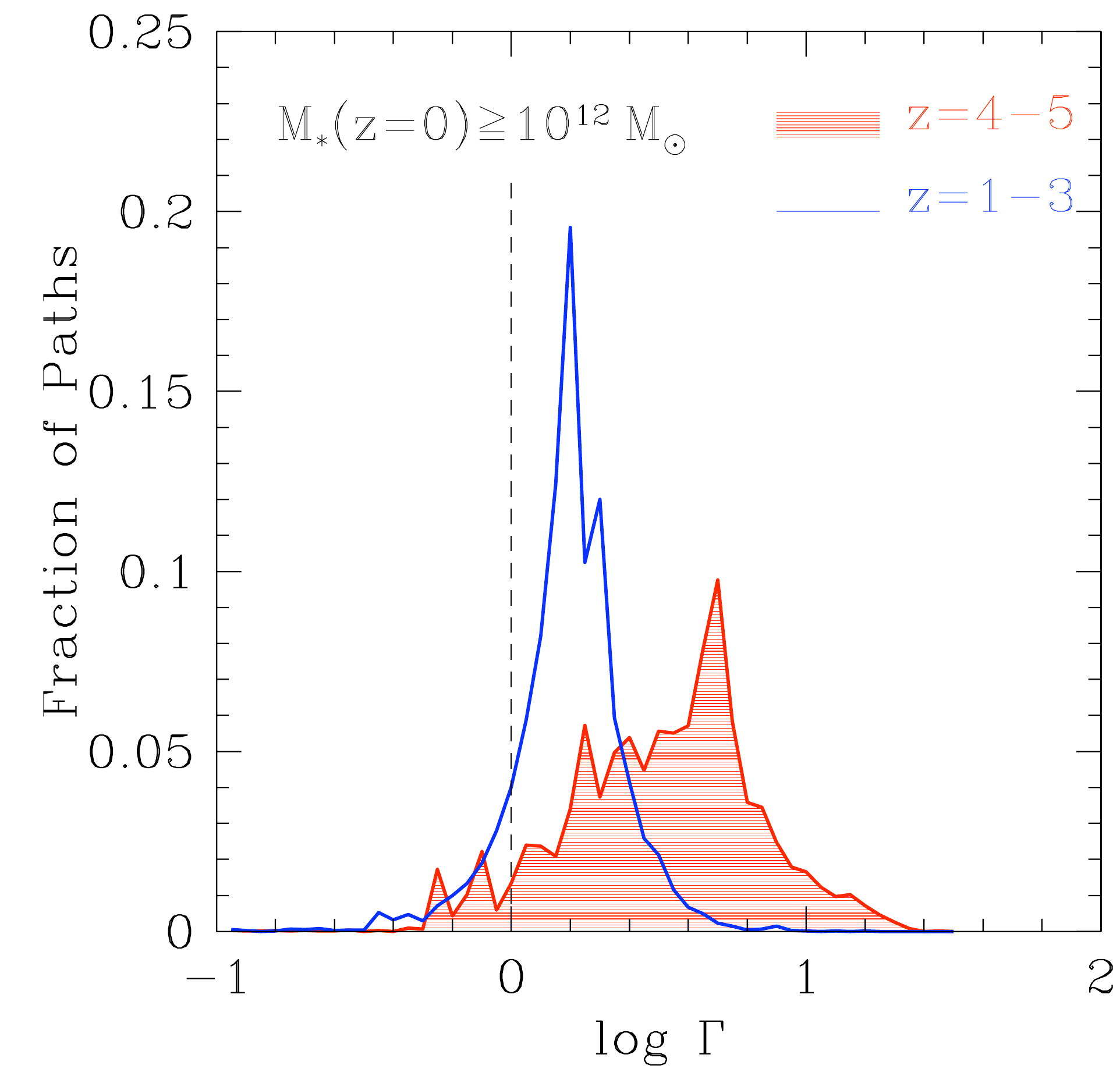}}}
\end{center}
\caption{Left Panel: The average values of $\Gamma$ at different redshifts (x-axis) corresponding to BH with a given final mass (y-axis) are represented as contours. The average is taken over all main progenitors in our Monte Carlo merging histories leading to 
the a given final mass.\newline
Right Panel: The fraction of paths leading to local massive galaxies with $M_*\geq 10^{12}\,M_{\odot}$ as a function of $log \Gamma$  is shown 
for two redshift bins: $4\leq z\leq 5$ (shaded region), and $1\leq z\leq 3$ (solid line). }
\end{figure*}

\begin{figure*}
\hspace{-2cm}
\vspace{0.1cm}
\scalebox{0.35}[0.35]{{\includegraphics{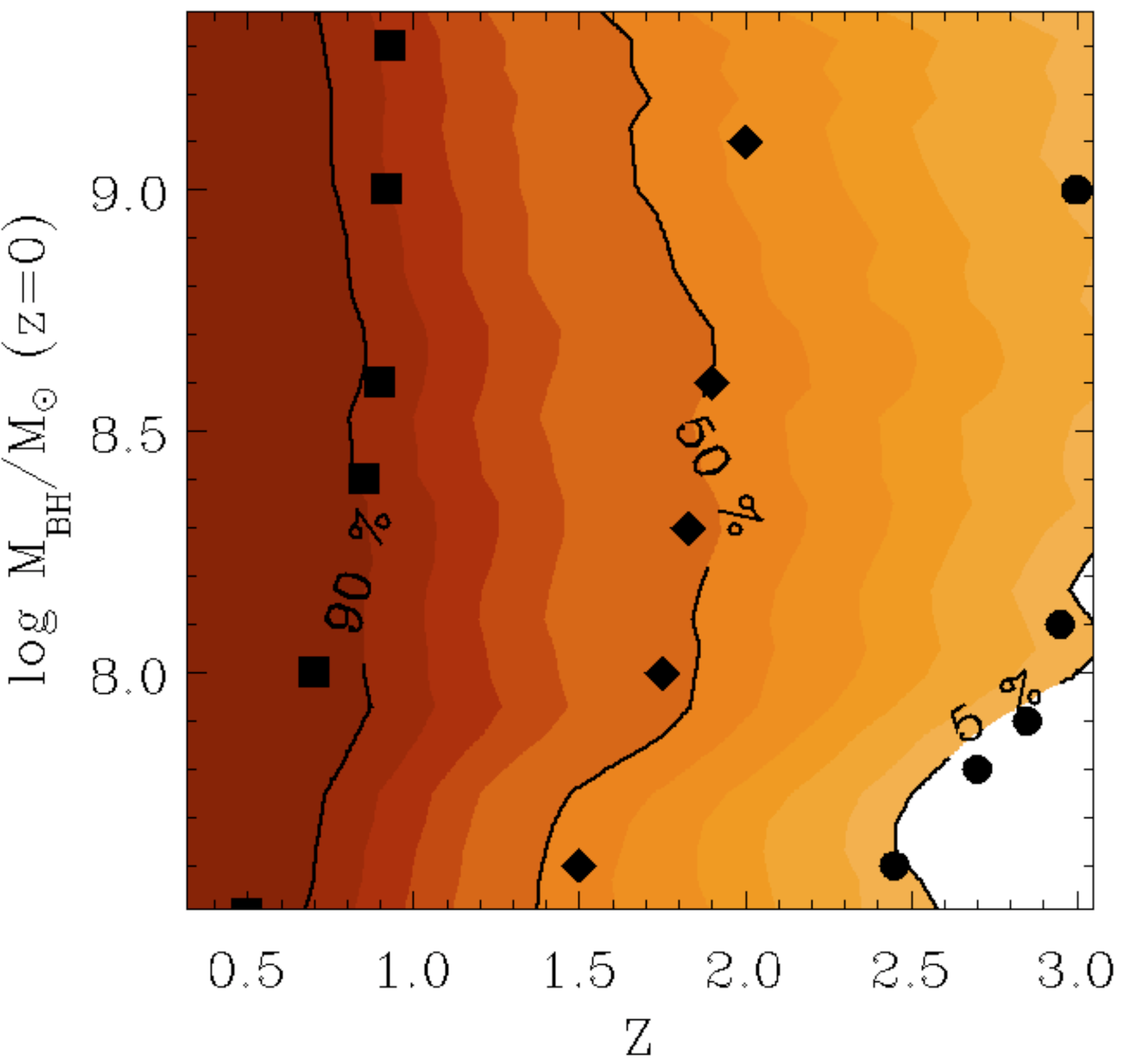}}}
\scalebox{0.35}[0.35]{{\includegraphics{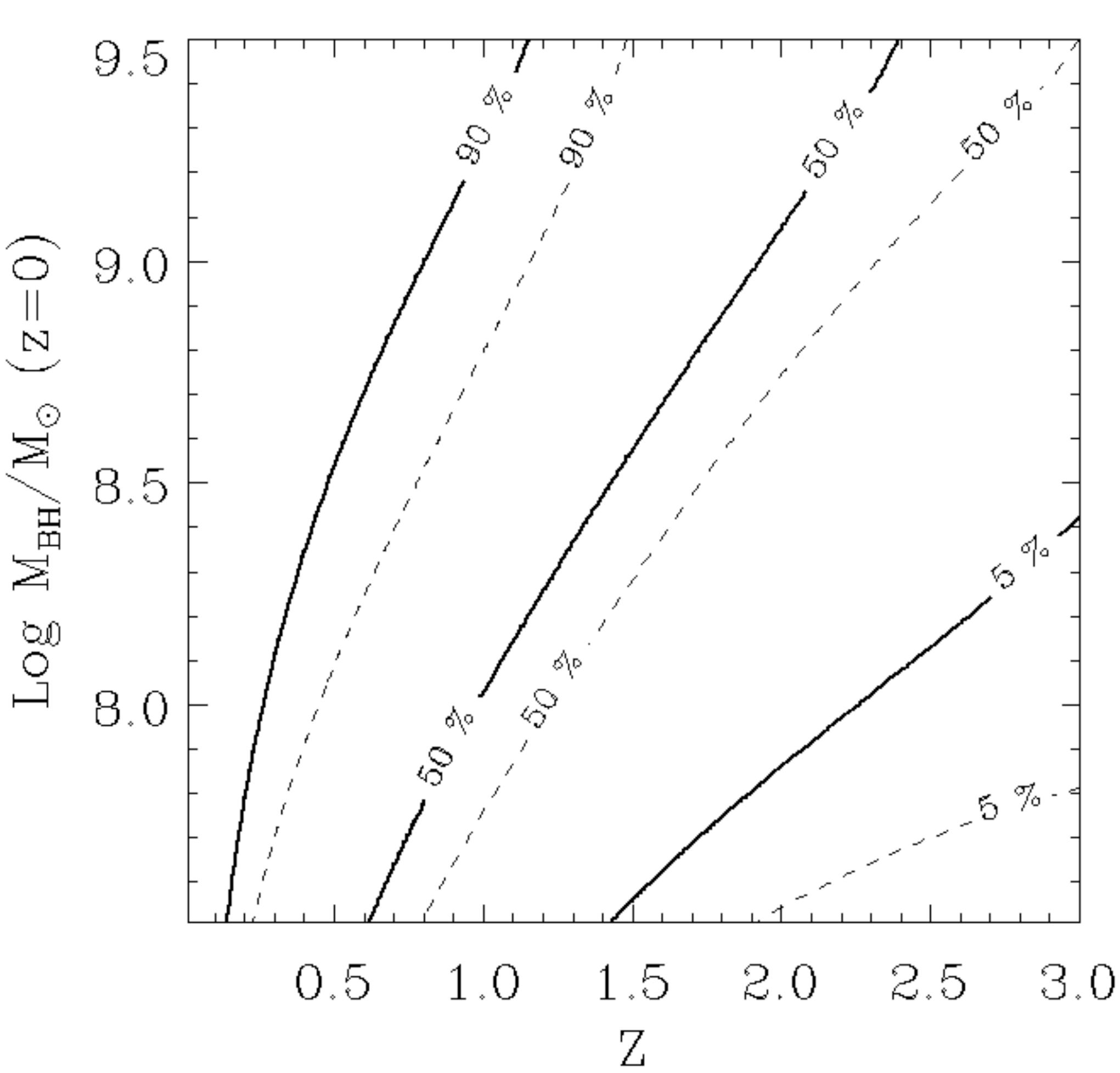}}}
\caption{Left Panel: The predicted growth histories of BH with different final mass. For
any final BH mass (represented in the y-axis), the 10 filled contours correspond
to different values of the average fractional mass $M_{BH}(z)/M_{BH}(z=0)$
of the progenitor BH: $M_{BH}(z)/M_{BH}(z=0)$=0.05, 0.1, 0.2, 0.3, 0.4, 0.5, 0.6, 0.7, 0.8, 0.9,
from  the lightest to the darkest contour. We have highlightened with thick black contours the levels
corresponding to values 0.9, 0.5, and 0.05, to compare with points obtained by
Marconi et al. (2004). These were obtained from the analysis of the evolution of
the observed AGN luminosity function (Ueda et al. 2003), under the assumption of
constant Eddington ratio $\lambda=1$. For such {\it observational} points, the
symbols squares, diamonds and circles, mark the contour levels corresponding to
the growth of 90 \%, 50\%, and 5\% of the final BH mass, respectively. \newline
Right Panel: We plot as continuous contours the same growth histories from the
analysis by Merloni (2009, see text), assuming that accretion at low $\dot m$ to
be radiatively inefficient (RIAF). The dashed contours show the same growth
history but assuming for all AGNs the same radiative efficiency (8 \%)
independently of $\lambda$.}
\end{figure*}

The above mass dependence of $\Gamma (z)$ has straightforward implications for the so-called AGN downsizing, i.e., the faster building-up of luminous AGNs compared with those with lower luminosities which is indicated by observations (see, e.g., Marconi et al. 2004); such a downsizing effect is also characterizing the faster drop (2.5 dex) from $z\approx 2$ to the present of luminous ($L_X\geq 10^{45}$ erg/s) AGNs compared to the low-luminosity population  ($L_X\leq 10^{43}$ erg/s), which is observed to follow only a gradual decrease (less than 1 dex in the number density) from 
$z=2$ to $z=0$ (see, e.g., Fiore et al. 2003; Hasinger , Miyaji \& Schmidt 2005; La Franca et al. 2005). In fact, the mass dependence of $\Gamma (z)$ represented in fig. 8 implies that the massive BH actually grow faster than the low-mass BHs. This is illustrated by fig. 9 (left panel), where the contours represent the fraction of BH mass formed at a given redshift (in the x-axis), for different values of the BH final mass in our model; 
any given fraction of the final mass is assembled at lower redshift for the less massive objects
 with $M_{BH}\lesssim 10^8\,M_{\odot}$ with respect to more massive objects.  
 We also reproduce as a sequence of points a similar contour plot produced by Marconi et al. (2004) based on the {\it observed} mass and luminosity evolution of AGNs, under the assumption of fixed Eddington ratio $\lambda=1$, and that galaxy merging does not affect the growth of SMBHs; although both assumptions  not hold in our model, it is interesting to note that the representation of downsizing naturally arising from the interaction-driven model resembles that extrapolated from observations.

The general observed trend of larger assembled fraction for increasingly massive BH is quite well accounted for by our interaction-driven model, 
although the model seems to show a bending in correspondence of the 50\% contour level at $z\approx 1.5$ for masses above $10^9$ $M_{\odot}$ which is not present in the data; this might constitute a true model inadequacy, or simply result from the simple assumptions of constant $\lambda$ adopted in the derivation of the data points by Marconi et al. (2004) from the analysis of the local BH mass function and of the 
evolution of the AGN luminosity functions. To clarify this issue we reproduce in fig. 9b the results from an improved version (Merloni 2009, in prep.) of the Merloni \& Heinz (2008) analysis, based on updated AGN luminosity functions. Such an analysis is similar in methodology to that by Marconi et al. (2004), but avoids the assumption $\lambda=1$ by deriving phenomenological, physically motivated relations between the accretion rate, the BH mass and the AGN luminosity,  and solving for a continuity equation for the SMBH mass function evolution.
Beyond the details concerning the exact values of the assembled mass fraction derived from the observed luminosity functions 
(which is still largely affected by uncertainties, as shown by the differences among the results obtained under different assumptions, see caption to fig. 9), the comparison with the Merloni \& Heinz (2008) approach shows i) that the BH growth inferred from the evolution of the AGN luminosity functions  generally yields a stronger downsizing effect at large masses  compared with our model, and ii) that such an effect is probably not due to the assumptions on $\lambda$ adopted to derive the BH growth histories from the observed luminosity functions. 
Thus, the discrepancy is likely to be due to the assumption (common to all the approaches based on the observed luminosity functions)  that the BH growth is always dominated by accretion, and that the BH growth due to merging is negligible at any $z$.

This conclusion is not unexpected when one considers the basic features of the hierarchical clustering picture. Indeed, in all approached based on hierarchical scenarios, the merging sector of the models predicts
massive objects to form later; such a trend is inverted when one consider how and when baryons are converted into stars or accreted into SMBHs. 
The overall  downsizing effect in hierarchical models thus results from the {\it balance} between the two effects above; the downsizing trend 
resulting at low-intermediate masses in our model means that accretion largely dominates over merging up to very large final BH masses 
$M_{BH}\geq 10^9\,M_{\odot}$.

While the comparison performed in Fig. 9 is just a first step toward a detailed observational testing of the downsizing properties of cosmological models for the growth of SMBHs, the global picture emerging from the interaction-driven model for the growth of SMBHs seems to consistently account for a wide set of independent observations concerning the growth of SMBHs; these range from the distant QSOs, to the intermediate-$z$ BL AGNs, to the extreme population constituted by the SMGs at $z\approx $2, to the average growth histories estimated from the observed luminosity functions, extending  over a wide range of redshifts $ 0\lesssim z\lesssim 4$ and of BH masses 
$10^7\lesssim  M_{BH}/M_{\odot}\lesssim10^9$.

\section{Conclusions}

We have investigated the growth of SMBHs relative to the stellar content of the host galaxy in the framework of an interaction-driven model for the feeding of BHs during the merging history of the host galaxies. Both the star formation and the BH accretion (i.e., the conversion of baryons from galactic gas into stars or into BHs) are ultimately determined by the histories of the host galaxy potential wells, strongly dependent on their total DM 
mass as predicted by galaxy formation models in a cosmological context. 
To test such a  model against observations,  we worked out specific predictions for sub samples of the simulated galaxy population corresponding to different observational samples for which observational estimates of 
black hole masses $M_{BH}$ and stellar masses $M_*$ were available. 
Specifically, for the evolution of the ratio $\Gamma\equiv 
(M_{BH}/M_*)(z)/(M_{BH}/M_*)(z=0)$ 
we found that: \newline
$\bullet$ Massive local galaxies ($M_*\geq 10^{12}\,M_{\odot}$) and their BHs have formed preferentially through paths (in the $M_*-M_{BH}$ plane) passing {\it above} the local  $M_*-M_{BH}$ relation.\newline
$\bullet$ The average value and the spread of $\Gamma(z)$ increase over the local value for increasing redshift (see fig. 7) . Such an increase is larger for massive BH (fig. 8), originating from biased overdense regions of the primordial density field. 
This constitutes a specific prediction of the interaction-driven 
models for the growth of SMBHs, since it is due to the effectiveness of interactions in triggering BH accretion in high-density environments at high redshifts $z\gtrsim 4$. In our model, the galaxy interactions constitute the only trigger for the BH growth, while star formation can proceed in a quiescent model at lower redshifts even in the absence of starbursts. Therefore, at high redshifts interactions rapidly trigger the accretion of BHs and the impulsive star formation, while at lower redshifts $z\lesssim 2$ the interaction rate drops, and only star formation can proceed in the quiescent model, leading to a decrease of $\Gamma$. The larger is the galaxy and BH mass, the faster is the gas conversion into stars and the BH growth at high redshift, the stronger the decline of $\Gamma$ at low $z$. \newline
$\bullet$ Due to the above physical mechanism, extremely massive BHs in  high redshift QSOs at $z\gtrsim 4$ are predicted to have typical values of $\Gamma\approx 4$ (see fig. 3), consistently with observational estimates (e.g., Willott et al. 2003; McLure \& Dunlop 2004; Vestergaard et al. 2004, Walter et al. 2004; Maiolino et al. 2007; Riechers et al. 2008); this corresponds to 
BHs growing faster than their host galaxy stellar mass in massive galaxies at high redshifts.  At lower $z$, lower mass BH hosted in luminous BL AGNs are predicted to have lower values of $\Gamma\approx 2$ (see fig.4). A direct implication of this is that AGN feedback begins to affect the galaxy properies already at high redshifts $z\gtrsim 4$.\newline
$\bullet$ In such a framework, SMG galaxies correspond to the rare merging histories leading to i) a large fraction of gas available for impulsive star formation at $z\approx 2$, ii) major merging events at such redshifts. These peculiar merging histories are characterized by lower interaction rates at high redshifts, which otherwise would have exhausted the galactic cold gas reservoirs; in turn, this corresponds to lower values of $\Gamma\approx 0.3-1$ for such objects (fig. 5), which are predicted to be building up a substantial fraction of their final BH mass at such redshifts $z\approx 2-3$. 
In addition, this objects are predicted to end up their evolution as low-to-intermediate mass BHs with $M_{BH}\lesssim 10^9\textbf{}$ $M_{\odot}$ (fig. 
6); such a prediction is in agreement with the independent findings of Alexander et al. (2008), 
based on the large number density of SMGs which is larger than that of local galaxies hosting BHs with masses exceeding $10^9$ 
$M_{\odot}$; in fact, the observed SMG density is consistent with the model predictions, as shown in Sect. 4.3.

The above comparison with observations probe the model in different range of masses and redshifts. The global picture emerging from the model 
(see Discussion  in Sect. 5) is consistent with a downsizing scenario, where massive BHs accrete a larger fraction of their final mass at high redshifts $z\geq 4$; this is a natural outcome of cosmological interaction-driven model for the growth of BHs, due to the larger effectiveness of interactions in high-density environments, which constitute the density regions where massive objects form. Note that this is entirely consistent with the hierarchical building up of cosmic structures, since the \textit{number density} of such massive objects is however increasing with time.

\section*{Acknowledgments} We thank A. Marconi for useful discussions and for his help in correcting the observed BH masses for the effect of radiation pressure. We acknowledge grants from ASI-INAF I/016/07/0.


\appendix

\section{The observed black hole--galaxy mass ratio at 3.9$<z<$6.4}

 A detailed discussion on the BH and galaxy masses in QSO at $z$ $\gtrsim$4 will be presented in a forthcoming paper by Riechers et al. (priv. comm.). 
In this appendix we obtain estimates of the BH and galaxy masses for a sample of four QSO at $z$ $\gtrsim$4 based on informations already available in the literature.
The $\Gamma$ values at 3.9$<$z$<$6.4 in Figs.1, 3 and 7, were obtained from
a few quasars whose dynamical mass of the host galaxy could be inferred
by high angular resolution CO millimetric
observations. In paricular the CO kinematical
data were obtained from Walter et al. (2004), for SDSS1148+5251 at z=6.42,
from Maiolino et al. (2007), for SDSS0338+0021 at z=5.03,
from Riechers et al. (2008), for BRI1335-0417 at z=4.407, and
from Riechers et al. (2009), for APM08279+5255 at z=3.911.
Note that for SDSS0338+0021 the dynamical mass is actually only
an upper limit, since the CO emission is not resolved.
We corrected the inferred dynamical masses for an average inclination
of the molecular disc of 15$^{\circ}$ in agreement with the bias
inferred by Carilli \& Wang (2006) for the inclination of the discs
in quasar hosts. We inferred the stellar mass by subtracting from the
dynamical mass the gas mass (as reported in the same references
reporting the CO observations) and the black hole mass discussed below.
Black hole masses were inferred for the same quasars by exploiting
virial estimators based on the width of the broad emission lines
and the continuum luminosity, calibrated on local AGNs
(e.g. Vestergaard \& Peterson 2006, Onken et al. 2004).
In particular black hole masses were
estimated (by using optical and near-IR spectra) by the following authors:
Barth, Martini \& Nelson (2003), for SDSS1148+5251, Dietrich \& Hamann (2004) 
for SDSS0338+0021 (note that the BH mass in Maiolino et al.
2007 was underestimated due to calibration problems),
Shields et al. (2006) for RI1335-0417, Riechers et al. (2009) for APM08279+5255.
 For what concerns the lensing factor of APM08279+5255, we adopted the value $\mu$=4 for the CO emission in the host galaxy, as inferred by  Riechers et al. (2009). 
As discussed by the same authors, the lensing factor of the nuclear emission, which is important to  estimate the BH mass, can be significantly larger, if the latter is close to the caustic.
To calculate the BH mass of this source we take $\mu_{nuc}$ $\sim$ 20, which is the the logarithmic average of the former value and the previous estimates based on the analysis of the nuclear emission lensed image
(Ibata et al. 1999, Egami et al. 2000, Mu{\~n}oz et al. 2001, Lewis et al. 2002).\\

The BH masses were corrected for the effect of
radiation pressure on the viral estimates, which is particularly
important for such powerful QSOs, by following the prescription
given in Marconi et al. (2008).

\section{The black hole and stellar mass measurements in individual SMG galaxies}

In figs. 5a and 5b the $\Gamma$ values of six X-ray obscured and three BL SMG
galaxies are reported.
The BH mass estimates of the BL SMGs are taken from Alexander et
al. (2008). They estimated M$_{BH}$ from the luminosity and line width of the
broad  H$\alpha$ line and the Greene \& Ho (2005) virial black hole mass estimator. We have assigned an
uncertainty of 0.5 dex to these estimates, equal to the 1$\sigma$
uncertainty of the virial relation adopted by the authors.\\
The BH masses of the X-ray obscured SMG are derived from Alexander et
al. (2005). They estimated M$_{BH}$ by converting the
X-ray luminosity in the (0.5-8) keV band to the AGN bolometric luminosity
(assuming a bolometric correction of 6$^{+12}_{-4}$ \%) under the assumption
that the accretion is Eddington limited. In fig. 5a the vertical bars correspond
to the M$_{BH}$ we obtain assuming Eddington ratio ($\lambda$=L$_{bol}$/L$_{Edd}$) of 0.1, 0.2 and 1.
 
The stellar masses, M$_*$, of the obscured SMGs and of one BL SMG 
(SMGJ123635+621424, star in fig. 5b ) are from Borys et al. (2005). They
estimated stellar masses using rest-frame UV/near-IR luminosities and a
mass-to-light ratio of 3.2 for these stellar population. The 
uncertainties in these M$_*$ estimates are from 0.01 dex to 0.23 dex (see Borys et al. 2005).  

The $\Gamma$ values for the other two BL SMG (SMGJ131222.3+423814 and
SMGJ163706.5+405313, diamonds in fig. 5b) are obtained using  dynamical
mass estimates, which are derived from CO line width measurements (Coppin et
al. 2008; Greve et al. 2005).
CO line widths can be converted into dynamical masses  assuming a
size and inclination for the gas reservoir. These estimates assumed a spatial
extent of the CO gas of R$\simeq$ 2 Kpc. The main uncertainties of these
dynamical mass estimates are due to the unknown extention of the CO region
and of the unknown inclination angle. It should also to be noted that
the dynamical mass includes  gas and stellar masses, therefore the
diamonds in fig. 5b represent lower limits for $\Gamma$.

\end{document}